\documentclass[12pt,aps,showpacs]{revtex4}
\usepackage{epsfig,amssymb,amsfonts,amsmath,bm}

\newcommand{\beq}{\begin{eqnarray}}
\newcommand{\eeq}{\end{eqnarray}}
\newcommand{\be}{\begin{equation}}
\newcommand{\ee}{\end{equation}}
\newcommand{\vex}{{\bm x}}
\newcommand{\ver}{{\bm r}}
\newcommand{\vesig}{{\bm \sigma}}
\newcommand{\vep}{{\bm p}}
\newcommand{\veS}{{\bm S}}
\newcommand{\veL}{{\bm L}}
\newcommand{\veD}{{\bm D}}
\newcommand{\veR}{{\bm R}}
\newcommand{\ven}{{\bm n}}
\newcommand{\veu}{{\bm u}}

\newcommand{\veB}{{\bm B}}
\newcommand{\veH}{{\bm H}}
\newcommand{\veE}{{\bm E}}

\newcommand{\velam}{{\bm \lambda}}
\newcommand{\llan}{\langle\langle}
\newcommand{\rran}{\rangle\rangle}
\newcommand{\lan}{\langle}
\newcommand{\ran}{\rangle}

\begin{document}

\title{Spin-dependent interactions in quarkonia}

\author{A. M. Badalian}
\affiliation{Institute of Theoretical and Experimental Physics, 117218, B.Cheremushkinskaya 25, Moscow, Russia}

\author{A. V. Nefediev}
\affiliation{Institute of Theoretical and Experimental Physics, 117218, B.Cheremushkinskaya 25, Moscow, Russia}

\author{Yu. A. Simonov}
\affiliation{Institute of Theoretical and Experimental Physics, 117218, B.Cheremushkinskaya 25, Moscow, Russia}

\begin{abstract}
The spin-dependent interactions in mesons are considered in detail in the framework of the Field Correlator Method. Analytic expressions for the
spin-dependent potentials in heavy and light quarkonia are derived with the QCD string moment of inertia taken into account. Recent lattice data
are analysed using these formulae and the data are shown to be consistent with very small values of the gluonic correlation length, $\lesssim 0.1$ fm.
The Field Correlator and the Eichten--Feinberg definitions of the spin-dependent potentials in the lowest, Gaussian approximation for the QCD vacuum
are compared to one another and the two approaches are shown to be equivalent in the limit of a vanishing vacuum correlation length, whereas for
finite values of the latter the difference between these two approaches can be explained by the contribution of the higher-order field corelators,
starting from the quartic one.
\end{abstract}
\pacs{12.38.Aw, 12.39.Ki, 12.39.Pn}

\maketitle
\section{Introduction}

Precise knowledge of spin-dependent quark--antiquark interactions
can be considered as one of the most important goals in the QCD
spectroscopy. It could help one to identify newly discovered orbital
excitations, calculate hadronic shifts of mesons with better
accuracy, to understand effects responsible for a suppression of
spin--orbital splittings in higher-mass mesons and baryons. Nevertheless, up to
now some important features of the spin-dependent interactions in
mesons remain unclear. In first turn it refers to the order
of levels inside $nP$ multiplets of heavy--light mesons and to small magnitudes of the fine
structure splittings of higher excitations.

It was assumed long ago that in heavy--light mesons the so-called
inversion of the $P$-wave levels could take place \cite{isg}. For
example, in the $D$- and $B$-mesons, the mass $M_2$ of the  $2^+$
($1^{3/2} P_2$) state might be smaller than the mass of the $0^+$
$(1^{1/2}P_0$) state. Here, in the notation $n^jP_J$, the symbol
$j$ refers to the total momentum of the light quark,
$j=l\pm\frac12$, whereas $n$, $l$, and $J$ stand for the radial
excitation number, the angular momentum ($l=1$), and the total
meson spin, respectively. The possibility of such an inversion was
checked by calculations in different models --- see, for example,
Refs.~\cite{Faust,KN}. Recently the absence of the spin--orbital
inversions in heavy--light mesons was studied in
Ref.~\cite{noinv} where, in a relativistic potential model, the
chiral radiative corrections were taken into account and it was
shown that such corrections provided the absence of the spin--orbital
inversion.

At present the most clean experimental data can be extracted from
the $D$- and $B$-mesons where the order of levels seems not to be
distorted by large hadronic shifts. All four members of the
$D(1P)$ multiplet are known \cite{PDG}: $D^*_0(2350)$,
$D_1(2422)$, $D_1(2427)$, and $D_2^* (2460)$. For the $D_{sJ}(1P)$
mesons two of them, $D^*_{s0} (2317)$ and $D_{s1}(2460)$, have
large hadronic shifts, $\gtrsim 100$ MeV \cite{DBs}, due to the
strong coupling to the $DK$ and $D^*K$ thresholds
\cite{ST1,8,R1}. Recently $B_1$, $B_2^*$ and $B_{s1}$,
$B^*_{s2}$ have been reported by the CDF and D\O\, collaborations
\cite{Bs}, for which $M_1<M_2$ (the notation $M_J$ is used for the
mass of the meson with the spin $J$; spin--orbital inversion
implies that $M_2<M_0$). Thus in the observed heavy--light mesons
the order of $1P_J$ levels is similar to that in heavy quarkonia,
where $M_0<M_1<M_2$, and coincides with the order specific for the
pure gluon--exchange interaction. In case of light and $K$-mesons
their experimental masses also satisfy the condition $M_0 <M_2$.

If one defines the ratio $\xi=a_{SO}/t$, where the matrix elements
$a_{SO}$ and $t$ are the spin--orbital and tensor splittings then,
for the Coulomb-type potential, $\xi_C=1.5$. Nonperturbative Thomas precession leads to a suppression of this ratio so that, in heavy quarkonia,
$\xi_{Q\bar Q} \lesssim 1.0$, and it is even smaller for the $D$-mesons,
$\xi_D\approx 0.8$. More specifically,
\begin{eqnarray*}
a_{SO}(\exp)&\simeq&1.04t(\exp)\quad(1P,b\bar{b}),\\
a_{SO}(\exp)&\simeq&1.03t(\exp)\quad(2P,b\bar{b}),\\
a_{SO}(\exp)&\simeq&0.86t(\exp)\quad(1P,c\bar{c}),\\
a_{SO}(\exp)&\simeq&0.77t(\exp)\quad(1P,D-{\rm meson}),
\end{eqnarray*}
that is, in all cases, the spin--orbital splitting is positive and it is either equal or not essentially smaller than the tensor splitting:
\be
0<a_{SO}(\exp)\lesssim t(\exp).
\label{4s}
\ee

In other words, the spin--orbital inversion is not observed in
experiment. This experimental fact contradicts the naive estimates
one can make in the framework of potential quark models. Indeed,
for spin-dependent interactions in heavy quarkonia (up to the
order $1/m^2$) the standard Eichten--Feinberg decomposition is
valid \cite{EF,12,EF2}:
\begin{eqnarray}
V_{SD}^{(0)}(r)&=&\left(\frac{\vesig_1\veL}{4\bar{m}_1^2}+\frac{\vesig_2\veL}{4\bar{m}_2^2}\right)\left[\frac1r\frac{dV_0}{dr}+\frac2r\frac{dV_1}{dr}\right]
+\frac{(\vesig_1+\vesig_2)\veL}{2\bar{m}_1\bar{m}_2}\frac1r\frac{dV_2}{dr}\nonumber\\
\label{SO0}\\
&+&\frac{(3(\vesig_1\ven)(\vesig_2\ven)-\vesig_1\vesig_2)}{12\bar{m}_1\bar{m}_2}V_3(r)+
\frac{\vesig_1\vesig_2 }{12\bar{m}_1\bar{m}_2}V_4(r)\nonumber,
\end{eqnarray}
where each potential $V_n(r)$ ($n=$0-4) contains both perturbative (P) and
nonperturbative (NP)
contributions: $V_n(r)=V_n^P(r)+V_n^{NP}(r)$.
The static interquark potential $V_0(r)$, together with the potentials $V_1(r)$ and $V_2(r)$, satisfies the Gromes relation \cite{Gr},
\be
V_0'(r)+V_1'(r)-V_2'(r)=0.
\label{Ge}
\ee
Notice that this relation refers both to the perturbative and nonperturbative parts of the potentials $V_n(r)$ $(n=0,1,2)$.

In the Eichten--Feinberg representation (\ref{SO0}) the quark masses $\bar{m}_1$ and $\bar{m}_2$
actually mean ``constituent masses'' which are not strictly defined. For example, even for the $b$-quark, $\bar{m}_b$ can vary
from the value $\sim 4.5$ GeV up to $5.2$ GeV in different potential models, while the pole mass of the $b$-quark,
$m_b(\mbox{2-loop})=4.79(8)$ GeV is known with the accuracy $\sim 80$ MeV \cite{PDG}. The situation is similar or even worse for lighter
quarks: the pole masses are $m_{u,d}(\mbox{2 GeV})\approx 3\div 8$ MeV, $m_s(\mbox{2 GeV})\approx 100(20)$ MeV,
$m_c\approx 1.40(7)$ GeV \cite{PDG}, while the ``constituent quark masses'' in  potential models take quite
different values, for example, $\bar{m}_{u,d}\cong 0.3$ GeV. As will be
shown below (see also Ref.~\cite{12}), well-defined average quark energies $\mu_i$ appear instead of $\bar{m}_i$
in the generalisation of Eq.~(\ref{SO0}).

Then, for a heavy--light meson with $m_1=m_q$ and $m_2=m_Q\gg
m_q$, the splittings $a_{SO}$ and $t$ are usually presented in the
form suggested in Ref.~\cite{CJ}:
\be
a_{SO}=\frac{1}{4\bar{m}_q^{2}}A+t,\quad t=\frac{1}{\bar{m}_q\bar{m}_Q}T,
\label{1s}
\ee
where in $a_{SO}$
the contribution proportional to $\bar m_Q^{-2}$ is neglected. For the ``linear+Coulomb" potential,
\be
V_0(r)=\sigma r-\frac43\frac{\alpha_S}{r}+\mbox{const},
\ee
and for the bound state with the angular momentum $l$ and the radial quantum number $n$, the factor
$A$ in the spin--orbital splitting (\ref{1s}) is therefore given by the matrix element ($V_1'(r)=-\sigma$ in this case --- see
Eq.~(\ref{V012eq}) below)
\be
A_{nl}=\lan\frac1r[V_0'(r)+2V_1'(r)]\ran_{nl}=\frac43 \alpha_S \lan r^{-3}\ran_{nl}-\sigma \lan r^{-1}\ran_{nl},
\label{3s}
\ee
which can become negative due to the second (Thomas precession) term in
Eq.~(\ref{3s}). Then, since the tensor splitting $t$ is inversely
proportional to the heavy-quark mass $\bar{m}_Q$ and is therefore
expected to be suppressed, one could naively expect that
$a_{SO}<0$ and the inversion of levels, $M_2<M_0$, has to take
place. In other words, the spin--orbital inversion could appear
due to a strong negative spin--orbital interaction which stems
from the confining linear potential.

More specifically, one can express the positions of the $0^+$ and $2^+$ levels via the matrix elements $a_{SO}$ and $t$ as follows \cite{QR1}:
\be
M(0^+)=M_0-2a_{SO}-t,\quad M(2^+)=M_0+a_{SO}-0.1t,
\ee
and hence the inversion of levels is possible if
\be
a_{SO}<0,\quad |a_{SO}|>0.3t.
\label{4a}
\ee
However, as was discussed before, the inequalities (\ref{4a}) do not take place
for the known heavy and heavy--light mesons.

As will be shown below, there are several reasons for that. First
of all, in Eq.~(\ref{3s}), the coefficient $A$ appears to be negative
and relatively small even for rather large values of the strong
coupling constant $\alpha_S$. Second, since the
first term in $a_{SO}$ is proportional to $\bar{m}_q^{-2}$, then,
as was already mentioned before, the correct definition of the
``mass'' $\bar{m}_q$ for the light and $s$ quarks is crucially
important. A model-independent definition of $\bar{m}_q$ as the
average kinetic energy of the quark is given in
Refs.~\cite{DKS,KNS}. It will be also given below. Furthermore,
the very mass factors in the spin-orbital terms in Eq.~(\ref{SO0})
are modified, if rotation of the string connecting the quarks and
the antiquark is taken into account. This type of corrections
gives a sizable effect in light and heavy--light mesons (in heavy
quarkonia the influence of string moment of inertia is much less
important) \cite{BNS} and suppresses $a_{SO}$ as a whole. As a
result, $a_{SO}$ appears to be positive and comparable with the
tensor splitting.

In other words, in order to explain experimental data one needs to
reconsider spin-dependent potentials in detail. This is the aim of
the present paper. We employ the Field Correlator Method (FCM)
\cite{FCM}. First, we use recent lattice data for spin--dependent
potentials in heavy quarkonia \cite{Komas} and extract the value
of the vacuum correlation length $T_g$ \cite{FCM}. We find that
the data are compatible with extremely small values of this
correlation length, less than 0.1~fm \cite{BNS2}. This value
appears to be even smaller than that estimated previously on the
lattice  \cite{Tg}. Notice that the physical role of $T_g$ for the
phenomenology of hadrons is quite important: in particular, for
hadrons of the spatial size $R$ and the temporal size $T_q$, the
QCD sum rule method can be applied if $R,T_q\ll T_g$, while
potential-type approaches are valid in the opposite limit,
$R,T_q\gg T_g$. With such a small value of the vacuum correlation
length we therefore justify the use of potential-type approaches
to quarkonia, in particular, the approach of the QCD string with
quarks at the ends, which will be used then in this paper.
Furthermore, the moment of inertia of the string between the
quark and the antiquark is taken into account, and we show how
this string moment of inertia affects spin--orbital splittings ---
it appears as an addendum to the quark energy in the denominators
in Eq.~(\ref{1s}) and thus it leads to the above mentioned
suppression of the term proportional to $A$.

For the sake of generality, wherever it is relevant, we keep the
contributions of the magnetic and electric correlators separately which allows us to comment
on briefly the deconfinement phase of QCD, where the  confining electric
correlators vanish and only the magnetic ones survive.

The paper is organised as follows. In Section~\ref{S1} we discuss spin-dependent interactions in heavy quarkonia and derive the
corresponding potentials in the framework of the FCM. We also discuss in detail the difference between the FCM and the Eichten--Feinberg definitions of
the spin-dependent potentials in heavy quarkonia. In particular, we demonstrate that this difference is due to the contribution of the higher-order field
correlators and that it is suppressed in the limit of the small vacuum correlation length. In Section~\ref{S2} we analyse the lattice data for the
spin-dependent potentials in heavy quarkonia and for the field strength Gaussian correlators and extract the vacuum correlation length.
We demonstrate the latter to be indeed small. Then, in Section~\ref{muvsm}, we generalise the form of the spin-dependent potentials (\ref{SO0})
for the case of light quarks defining the mass parameters $\bar{m}_i$ through the averaged kinetic energies of the quarks and taking into account the
QCD string moment of inertia. We conclude and discuss the results in Section~\ref{S4}.

\section{Spin-dependent potentials in quarkonia}\label{S1}

\subsection{Spin-dependent potentials in the FCM}

We start with the meson wave function in the in- and out-states
\cite{ST1,12,ST2},
\be
\Psi^{({\rm in, out})}_{q\bar q}(x,y|A)=\bar{\Psi}_{\bar q}(x)\Gamma\Phi(x,y)\Psi_q(y),
\ee
written in a gauge-invariant way with the help of the parallel
transporter,
\be
\Phi(x,y)=P\exp{\left(ig\int_{y}^{x}dz_{\mu}A_{\mu}^at^a\right)}.
\label{partr}
\ee
The matrix $\Gamma$ stands for the vertex
function and provides the correct quantum numbers of the meson.
Then the Green's function can be constructed as
\begin{eqnarray*}
G_{q\bar q}&=&
\langle\Psi_{q\bar q}^{({\rm out})}(x_2,y_2|A)
\Psi^{({\rm in})\dagger}_{q\bar q}(x_1,y_1|A)\rangle_{q\bar{q}A}\\[2mm]
&=&\langle {\rm Tr}S_q(x_2,x_1|A)\Gamma\Phi(x_1,y_1)S_{\bar{q}}(y_1,y_2|A)\Gamma^\dagger\Phi(y_2,x_2)\rangle_A,
\end{eqnarray*}
where $S_{q}$ and $S_{\bar{q}}$ are the propagators of the quark
and the antiquark, respectively, in the background gluonic field
and we discarded  the disconnected contribution. One can proceed
then by using the Fock--Feynman--Schwinger representation
\cite{FSr,Sim2,Lisbon} for the single-quark propagators, thus
arriving at
\be
G_{q\bar q}=\int_0^\infty ds_1 \int_0^\infty ds_2
Dz_1 Dz_2\; e^{-K_1-K_2}\lan {\rm Tr}
(m_1-\hat{D}_1)\Gamma(m_2-\hat{\bar{D}}_2^*)\Gamma^\dagger
W(C)\ran,
\label{6s}
\ee
where $K_{1,2}$ are the quark and
antiquark kinetic energy terms,
\be
K_1=\int_0^{s_1}\left[m_q^2+\frac14\dot{z}_1^2(\tau_1)\right]d\tau_1,\quad
K_2=\int_0^{s_2}\left[m_Q^2+\frac14\dot{z}_2^2(\tau_2)\right]d\tau_2,
\label{7s}
\ee
and the spin-independent interquark interaction is
described in terms of the Wilson loop $W(C)$, with the contour $C$
running over the quark trajectories:
\be
W(C)=\exp ig\int ds_{\mu\nu}(z)F_{\mu\nu}(z),
\label{trw}
\ee
with $s_{\mu\nu}$ being the surface element. In Eq.~(\ref{7s}), $m_q$ and $m_Q$ are
the conventional pole masses of the quark and the antiquark,
whereas $s_{1,2}$ and $\tau_{1,2}$ are the proper time variables,
introduced by Fock and Schwinger \cite{FSr}; $z_{1\mu}(\tau_1)$
and $z_{2\mu}(\tau_2)$ are the paths of the quark and antiquark,
respectively.

The pre-exponential term in Eq.~(\ref{6s}) stems from the fermionic Dirac projectors. It can be expressed through the
derivatives with respect to the surface element and thus it can be pulled out from the average over the background gluonic field.
This term contributes to the spin-dependent interquark interactions \cite{Sim2}
(see Appendices~\ref{colel} and \ref{colmagn} for the details).
The most economical way to include the spin-dependent terms in the exponent in Eq.~(\ref{6s}) is to extend the
differential in Eq.~(\ref{trw}) to include the spinor structure
\cite{Lisbon}:
\be
ds_{\mu\nu}(z)\to d\pi_{\mu\nu}(z)=ds_{\mu\nu}(z)-i\sigma^{(1)}_{\mu\nu}d\tau_1+i\sigma_{\mu\nu}^{(2)}d\tau_2,
\ee
with $\sigma_{\mu\nu}^{(i)}=\frac{1}{4i}(\gamma_\mu\gamma_\nu-\gamma_\nu\gamma_\mu)$ ($i=1,2$ for the quark and antiquark, respectively).
Then the spin-dependent interactions result from the mixed terms containing the spin variables.

The averaged Wilson loop (\ref{trw}) can be expressed through the correlators of the field strength tensors as
\begin{eqnarray}
\langle{\rm Tr}W(C)\rangle&=&\langle{\rm Tr}\exp ig\int d\pi_{\mu\nu}(z)F_{\mu\nu}(z)\rangle\nonumber\\[-2mm]
\label{W1}\\[-2mm]
=\exp\sum^\infty_{n=1}\frac{(ig)^n}{n!}&\displaystyle\int& d\pi(1)
\ldots\int d \pi(n)\langle\langle F(1)\ldots F(n)\rangle\rangle\nonumber,
\end{eqnarray}
where the cluster expansion theorem  was used (see  Ref.~\cite{FCM} for the relevant
references and for the detailed discussion). The average
$\langle\langle\ldots\rangle\rangle$ stands for connected
correlators, for example, for the bilocal correlator,
$\langle\langle F(1)F(2)\rangle\rangle=\langle
F(1)F(2)\rangle-\langle F(1)\rangle\langle F(2)\rangle$, and
$F_{\mu\nu}=\partial_\mu A_\nu-\partial_\nu A_\mu- ig [A_\mu,
A_\nu]$ is the vacuum field strength. Obviously, due to the $O(4)$
rotational invariance and colour neutrality of the vacuum,
$\langle\langle F\rangle\rangle=\langle F\rangle=0$.

In the Gaussian approximation for the vacuum, when only the
lowest, bilocal correlator is retained one has, with the accuracy of a few per cent
(see Ref.~\cite{Cas} for the discussion):
\be
\langle Tr
W(C)\rangle\propto\exp\left[-\frac12\int_Sd\pi_{\mu\nu}(x)d\pi_{\lambda\rho}(x')
D_{\mu\nu\lambda\rho}(x-x')\right], \label{W2} \ee where
$$
D_{\mu\nu\lambda\rho}(x-x')\equiv\frac{g^2}{N_c}\langle\langle{\rm Tr}F_{\mu\nu}(x)\Phi(x,x')F_{\lambda\rho}(x')\Phi(x',x)\rangle\rangle.
$$
This bilocal correlator of gluonic fields can be expressed through only two gauge-invariant scalar functions $D(u)$ and $D_1(u)$ as \cite{FCM}
\be
D_{\mu\nu\lambda\rho}(u)=(\delta_{\mu\lambda}\delta_{\nu\rho}-\delta_{\mu\rho}\delta_{\nu\lambda})D(u)
+\frac12\left[\frac{\partial}{\partial u_\mu}(u_\lambda\delta_{\nu\rho}-u_\rho\delta_{\lambda\nu})+
\genfrac{(}{)}{0pt}{0}{\mu\leftrightarrow\nu}{\lambda\leftrightarrow\rho}\right]D_1(u).
\label{Dcordef}
\ee
The correlator $D(u)=D(u_0,|\veu|)$ contains only a nonperturbative part and it is responsible for the QCD string formation at large interquark
separations. The fundamental string tension can be expressed as a double integral:
\be
\sigma=2\int_0^\infty d\nu\int_0^\infty d\lambda D(\nu,\lambda).
\label{sigma}
\ee

The spin--dependent terms in the interquark interaction are generated by
the combination $\sigma_{\mu\nu}F_{\mu\nu} $ present in Eq.~(\ref{W1}), which
reads: \be \sigma_{\mu\nu}F_{\mu\nu}= \left(
\begin{array}{ll}
\vesig \veH& \vesig\veE\\
\vesig\veE&\vesig \veH
\end{array}
\right),
\label{3}
\ee
and therefore one needs correlators of the colour-electric and colour-magnetic fields, as well as mixed terms, separately. They immediately follow from
the general expression (\ref{Dcordef}) and read \cite{FCM}:
\beq
\frac{g^2}{N_c}\lan\lan {\rm Tr} E_i(x)\Phi E_j(y)\Phi^\dagger\ran\ran&=&\delta_{ij}\left(D^E(u)+D_1^E(u)+u^2_4\frac{\partial D_1^E}{\partial u^2}\right)+
u_iu_j\frac{\partial D_1^E}{\partial u^2},\\
\frac{g^2}{N_c}\lan\lan {\rm Tr} H_i(x)\Phi H_j(y)\Phi^\dagger\ran\ran&=&\delta_{ij}\left(D^H(u)+D_1^H(u)+\veu^2\frac{\partial D_1^H}{\partial\veu^2}\right)-
u_iu_j\frac{\partial D_1^H}{\partial u^2},\label{Hs0}\\
\frac{g^2}{N_c}\lan\lan {\rm Tr} H_i(x)\Phi
E_j(y)\Phi^\dagger\ran\ran&=&\varepsilon_{ijk}
u_4u_k\frac{\partial D_1^{EH}}{\partial u^2},
\label{HE0}
\eeq
where $u_\mu=x_\mu-y_\mu$, $u^2=u_\mu u_\mu$. We keep here the
superscripts $E$ and $H$ in the correlators $D$ and $D_1$ in order
to distinguish the electric and magnetic parts of the correlators
and thus to be able to consider a nonzero temperature $T$. Indeed,
while $D^E=D^H$ and $D_1^E=D_1^H=D_1^{EH}$ at $T=0$, at higher
temperatures they behave differently. In particular, above the
deconfinement temperature, $T>T_c$, the electric correlator $D^E$
disappears, whereas $D_1^E$ and the magnetic correlators survive.

The spin-independent interquark interaction is generated by the
term $\propto ds_{\mu\nu}(x) ds_{\lambda\rho}(x')$ in the Wilson
loop (\ref{W2}). It is given by the standard Nambu--Goto action
for the minimal surface area,
\be
A_{\rm string}=\sigma S_{\rm min},\quad S_{\rm min}=\int_0^Tdt\int_0^1d\beta\sqrt{(\dot{w}w')^2-\dot{w}^2w'^2},
\label{Smin}
\ee
where for the profile function of the string
$w_\mu(t,\beta)$ we adopt the straight--line ansatz: \be
w_{\mu}(t,\beta)=\beta x_{1\mu}(t)+(1-\beta)x_{2\mu}, \label{anz}
\ee $x_{1,2}(t)$ being the four--coordinates of the quarks at the
ends of the string. This approximation is valid at least for not
high excitations due to the fact that hybrid excitations
responsible for the string deformation are decoupled from a meson
by the mass gap of order 1~GeV.

It is straightforward now to derive the spin-dependent
interactions in heavy quarkonia in the framework of the FCM. For
example, the spin--orbital interaction results from the mixed
terms $ds_{\mu\nu}\sigma_{\lambda\sigma}d\tau$ in (\ref{W2}) and
reads:
\be
L_{SO}=\int d
s_{\mu\nu}(w)d\tau_1\sigma^{(1)}_{\lambda\rho}D_{\mu\nu\lambda\rho}(w-x_1)+(1\to 2),
\label{LSO26}
\ee
where
\be
ds_{\mu\nu}=\varepsilon^{ab}\partial_a w_\mu(t,\beta)\partial_b
w_\nu(t,\beta)dtd\beta,\quad a,b=\{t,\beta\}.
\ee
For the straight-line string ansatz (\ref{anz}) this yields:
\be
ds_{i4}=r_idtd\beta,\quad
ds_{ik}=\varepsilon_{ikm}\rho_mdtd\beta,\quad {\bm \rho}=[\ver\times(\beta\dot{\vex}_1+(1-\beta)\dot{\vex}_2)],
\label{19a}
\ee
and thus the angular momentum appears from $ds_{ik}$. Further details of the derivation, as well as the
comparison with the lattice data, can be found in Refs.~\cite{SO2,Lisbon}. The
result can be presented in the form of Eq.~(\ref{SO0}) with the following identification of the potentials
($D(u)=D(\lambda,\nu)$, $D_1(u)=D_1(\lambda,\nu)$,
$u=\sqrt{\lambda^2+\nu^2}$):
\begin{eqnarray}
V_0'(r)&=&2\int_0^\infty d\nu\int_0^rd\lambda D^E(\lambda,\nu)+r\int_0^\infty d\nu D_1^E(r,\nu),\nonumber\\
V_1'(r)&=&-2\int_0^\infty d\nu\int_0^r d\lambda \left(1-\frac{\lambda}{r}\right)D^H(\lambda,\nu),\nonumber\\
V_2'(r)&=&\frac{2}{r}\int_0^\infty d\nu\int_0^r \lambda d\lambda D^H(\lambda,\nu)+r\int_0^\infty d\nu D_1^H(r,\nu),\label{Vseq}\\
V_3(r)&=&-2r^2\frac{\partial}{\partial r^2}\int_0^\infty d\nu D_1^H(r,\nu),\nonumber\\
V_4(r)&=&6\int_0^\infty d\nu\left[D^H(r,\nu)+\left[1+\frac23r^2\frac{\partial}{\partial\nu^2}\right]D_1^H(r,\nu)\right].\nonumber
\end{eqnarray}
and with the masses $\bar{m}_i$ replaced by the dynamically generated effective quark masses $\mu_i$, corrected to account for the inertia of the
QCD string --- see Section~\ref{muvsm} below for a detailed discussion.
In Appendices~\ref{colel} and \ref{colmagn} we give the derivation of the terms $V_0'$,
$V_1'$, and $V'_2$ via field correlators with the string moment of inertia taken into account.

With the explicit form of the potentials (\ref{Vseq}) found in the framework o the FCM one can check the Gromes relation (\ref{Ge}), which now reads:
\begin{eqnarray}
V_0'(r)+V_1'(r)-V_2'(r)=2\int_0^\infty d\nu\int_0^r d\lambda [D^E(\lambda,\nu)-D^H(\lambda,\nu)]\nonumber\\[-3mm]
\label{Ge2}\\[-3mm]
+r\int_0^\infty d\nu[D_1^E(r,\nu)-D_1^H(r,\nu)]\nonumber.
\label{20s}
\end{eqnarray}

The right-hand side of Eq.~(\ref{20s}) vanishes at the temperature $T=0$, when $D^E(u)=D^H(u)$ and $D_1^E(u)=D_1^H(u)$
\cite{DEH}. Notice that, in the FCM, the Gromes relation holds both for the perturbative and nonperturbative parts of the potentials (\ref{Vseq}).
For $T\neq 0$, in general case, $D^E\neq D^H$ and the Gromes relation does not hold since $O(4)$ (Lorentz) invariance is
violated. Therefore lattice measurements of the Gromes relation at finite temperatures, especially in the deconfinement phase of QCD, could shed new light on
the nature of the vacuum fields.

The correlator $D(u)$ does not contain perturbative part \cite{SS1} (note that, for this reason, $V'_1(r)$ --- see
Eq.~(\ref{Vseq}) --- does not contain perturbative contributions either) and decreases in all directions of the Euclidean space.
This decrease is governed by the gluonic vacuum correlation length $T_g$ and has an exponential form, as was found analytically
in Ref.~\cite{simcor} and measured on the lattice \cite{Tg}. Then, this correlator has the form:
\be
D(u)=\frac{\sigma}{\pi T_g^2}\exp[-u/T_g],
\label{Dcor0}
\ee
where the coefficient is chosen to satisfy the relation (\ref{sigma}). The form (\ref{Dcor0}) is oversimplified.
In particular, it is not regular at $u=0$, while a more suitable form should behave as $\propto u^2$ at small distances $|u|<T_g$
and then it exponentially decreases at large $u$'s ---
this follows from the gluelump Green's function studied in Ref.~\cite{simcor}. However, this small change with respect to the form (\ref{Dcor0})
is not felt in the integrals of $V_n(r)$ and cannot be noticed in comparison of the field correlators with the lattice simulations,
since the region $|u|<T_g$ is not measured on the lattice. The regular Gaussian form was used in Ref.~\cite{SO3} and both forms were shown
to yield similar results.

Another important comment concerning the correlation length $T_g$ is the scale at which it is defined (see, for example, Ref.~\cite{BM1} for the
discussion of the issue in relation to the Operator Product Expansion in QCD).
By natural arguments we expect the scale of $T_g$ to be
of order of the average size of the gluelump (average momentum), which is of
the order of 1 GeV (see Ref.~\cite{gluelump}).

The other correlator, $D_1(u)$, contains both
perturbative and nonperturbative contributions. Its perturbative
part leads to the colour Coulomb interaction between quarks,
whereas its nonpeturbative part was found through the gluelump
Green's function in Ref.~\cite{simcor}:
\be
D_1(u)=D_1^P(u)+D_1^{NP}(u)=\frac{16\alpha_S}{3\pi}\frac{1}{u^4}+\frac{\sigma_1}{\pi T_g'u}\exp[-u/T_g'],
\label{D1cor0}
\ee
where, similarly to Eq.~(\ref{sigma}), we have defined
\be
\sigma_1\equiv 2\int_0^\infty d\nu\int_0^\infty d\lambda D_1^{NP}(\nu,\lambda)=6\pi\alpha_S\sigma,\quad\alpha_S=\frac{g^2}{4\pi}.
\label{sigma1}
\ee
Notice that, for the sake of generality, we consider two
correlation lengths, $T_g$ and $T_g'$, for the correlators $D(u)$
and $D_1^{NP}(u)$, respectively (see Ref.~\cite{simcor} for the
details) and we find them to be different \cite{BNS2}.

For large interquark separations, $r\gg T_g,T_g'$, one can find the following asymptotic formulae for the potentials (\ref{Vseq}):
\be
V_0'(r)=\sigma+\frac43\frac{\alpha_S}{r^2},\; V_1'(r)=-\sigma,\; V_2'(r)=\frac43\frac{\alpha_S}{r^2},\;
V_3(r)=\frac{4\alpha_S}{r^3},\; V_4(r)=\frac{32}{3}\pi\alpha_S\delta^{(3)}(r).
\label{V012eq}
\ee
In particular, the static interquark potential comes out from Eq.~(\ref{V012eq}) in the standard ``linear+Coulomb" form,
$$
V_0(r)=V_{Q\bar{Q}}(r)=\sigma r-\frac43\frac{\alpha_S}{r}+{\rm const}.
$$

\subsection{Eichten--Feinberg versus FCM definition of the potentials
$V_1'(r)$ and $V_2'(r)$}\label{FCMvsEF}

\begin{figure}[t]
\begin{center}
\epsfig{file=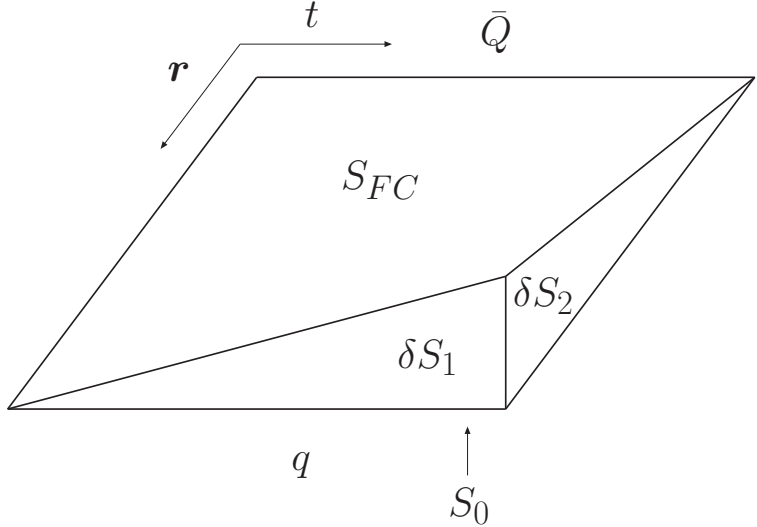,width=10cm}
\caption{The surfaces $S_{FC}$ and $S_{EF}=S_0+\delta S_1+\delta S_2$.}\label{surfsfig}
\end{center}
\end{figure}

In this chapter we compare the definition of the potentials $V_1'(r)$ and $V_2'(r)$ in the FCM, as given in Eq.~(\ref{Vseq}),
with the Eichten--Feinberg (EF) definition,
\beq
\frac{r_k}{r}\frac{dV^{\rm EF}_1(r)}{dr}=\varepsilon_{ijk}\lim_{T\to\infty}&\displaystyle\int_0^T dt&t\llan g^2 H_i(\ver_1,t_1)E_j(\ver_1,t_2)\rran,\label{C1}\\
\frac{r_k}{r}\frac{dV^{\rm EF}_2(r)}{dr}=\varepsilon_{ijk}\lim_{T\to\infty}&\displaystyle\int_0^T dt&t\llan g^2 H_i(\ver_1,t_1)E_j(\ver_2,t_2)\rran,\label{C2}\\
\left(\frac{r_ir_j}{r^2}-\frac{\delta_{ij}}{3}\right)V_3^{\rm EF}(r)+\frac{\delta_{ij}}{3}V_4^{\rm EF}(r)=
\lim_{T\to\infty}&\displaystyle\int_0^T dt&\llan g^2 H_i(\ver_1,t_1)H_j(\ver_2,t_2)\rran,\label{C3}
\eeq
with $t=t_2-t_1$. The average is introduced as
\be
\llan O(1)O(2)\rran={\rm Tr}\lan O(1)O(2)W(S_0)\ran,
\label{C333}
\ee
with $O(i)$ ($i=1,2$) meaning the insertion of a field operator (plaquette) in the unperturbed rectangular Wilson loop $W(S_0)$.
For the sake of simplicity we consider the limit of the antiquark being infinitely heavy, $m_{\bar{Q}}=m_2=\infty$.

Let us start from the quark--antiquark Green's function (\ref{6s}) and notice that the magnetic part of the spin--orbital interaction
appears from the average (see Refs.~\cite{SO2,SO3} for the details):
\be
\lan gH_i (\ver_1,t_1)W(C)\ran.
\label{C4}
\ee
In the last equation, the Wilson loop $W(C)$ is taken along the contour formed by the trajectories of the quark and the antiquark,
and the averaging is assumed over all such trajectories as well as over all vacuum field configurations.
For heavy quarks, the deflection of the perturbed path $C$ from the straight-line path $C_0$ is small (see Fig.~\ref{surfsfig}),
so that one can envisage a sort
of perturbative expansion in powers of this small deflection. Below we shall perform such an expansion in two ways: 1) in the manner of the
Eichten--Feinberg approach; 2) in the framework of the FCM. In the given two approaches one uses different choices of the
surface bounded by the contour $C$. Indeed, in the FCM this is the minimal surface, $S_{FC}=S_{\rm min}(C)$,
approximated by the straight-line ansatz --- see Eq.~(\ref{anz}) ---
while, in the EF approach, the surface is given by the sum of three pieces (see Fig.~\ref{surfsfig}),
\be
S_{EF}(C)=S_0+\delta S_1+\delta S_2.
\label{C5}
\ee

Due to the nonabelian Stokes theorem, the Wilson loop average (\ref{C4}) does not depend
on the choice of the surface, hence
\be
\lan g H_i(\ver_1, t_1)W(S_{EF})\rangle=\lan g H_i(\ver_1,t_1)W(S_{FC})\rangle.
\label{C7}
\ee
One can conclude therefore that the two approaches are equivalent, provided {\em the full infinite set of field correlators is taken into account}.
In the meantime, in both approaches, only the leading correlators are retained, and this leads to a discrepancy between the definitions of the
potentials discussed before. Below we explain this in more detail.

In the FCM one simply uses the cluster expansion for the
right-hand side of Eq.~(\ref{C7}) and, keeping only bilocal correlators, arrives at
(see Refs.~\cite{SO3} for the details):
\be
\lan g H_i(\ver_1, t_1) W(S_{FC})\ran\approx i\int ds_{\mu\nu}(w)\lan g^2 H_i(\ver_1,t_1)F_{\mu\nu}(w)\ran.
\label{C8}
\ee
Then, following the same steps as in Appendix~\ref{colmagn}, it is easy to arrive at the formula (notice that, for the sake of simplicity,
we omit here the string inertia and consider heavy quarks):
\be
\frac{dV_1^{FC}(r)}{dr}=-2\int_0^\infty d\nu\int_0^r d\lambda\left(1-\frac{\lambda}{r}\right)D^H(\lambda,\nu),
\label{C9}
\ee
which coincides with that given in Eq.~(\ref{Vseq}). This result is accurate up to the quartic correlators $\llan F^4\rran$ and higher.

Let us now turn to the EF case --- the left-hand side of Eq.~(\ref{C7}). Notice that the pieces $S_0$ and $\delta S_2$ (see Eq.~(\ref{C5})
and Fig.~\ref{surfsfig}) do not contribute to the average. Indeed, in the sum
\beq
\lan gH_i(\ver_1, t_1)W(S_{EF})\ran=ig^2\sum_{S=S_0,\delta S_1,\delta S_2}\int_Sds_{\mu\nu}\lan H_i(\ver_1,t_1)F_{\mu\nu}(w) W(S_{EF})\ran\nonumber\\
=i\int_{S_0}ds_{\mu\nu}\lan g^2H_i(\ver_1,t_1)F_{\mu\nu}(w)\ran+i\int_{\delta S_2}ds_{\mu\nu}\lan g^2H_i(\ver_1,t_1)F_{\mu\nu}(w)\ran\\
+i\int_{\delta S_1}ds_{\mu\nu}\lan g^2H_i(\ver_1,t_1)F_{\mu\nu}(w)W(S_{EF})\ran\nonumber
\eeq
only the last term contributes, since the integral over $S_0$ is odd with respect to the inversion of one of the spatial coordinates, whereas
the integral over $\delta S_2$ vanishes for $t_1\gg T_g$. The remaining integral over the surface $\delta S_1$ can be written as
\be
\lan gH_i(\ver_1, t_1)W(S_{EF})\ran=i\lan g^2 H_i(\ver_1, t_1)\int_{\delta S_1}ds_{j4} E_j(u,t_2)W(S_0)\rangle,
\label{C11}
\ee
with
\be
W(S_0)=\exp ig \int_{S_0}ds_{n4}(w)E_n(w),
\ee
and it is easy to check that its cluster expansion starts with the triple correlator with the leading correction coming from the fifth-order
correlators $\llan F^5\rran$.

In order to proceed we write
\begin{eqnarray*}
&&i\lan g^2 H_i(\ver_1,t_1)E_j(\ver_1, t_2)W(S_0)\ran\equiv \varepsilon_{ijn}r_n\times\frac12
\varepsilon_{ijk} \frac{r_k}{r^2} i\lan g^2 H_i(\ver_1,t_1)E_j(\ver_1, t_2)W(S_0)\ran,\\
&&ds_{j4}=(x_{1j}(t_2)-r_{1j})dt_2\approx\dot x_{1j}(t_2)(t_2-t_1)dt_2,\\
&&\varepsilon_{ijn}r_n\dot x_{1j}(t_2)=[\ver\times\dot{\vex}_{1i}]=\frac{L_{1i}}{im_1},
\end{eqnarray*}
where $\veL_1$ is the quark angular momentum. Then
\begin{eqnarray*}
\Delta L_{SO}^H&=&-i\int d\tau_1 \vesig_1 \mbox{Tr}\lan g \veH(\ver_1, t_1) W(S_{EF})\ran\\
&=&\int\frac{dt}{2m_1}ds_{j4}\sigma_{1i}\mbox{Tr}\lan g^2 H_i(\ver_1,t_1)E_j(\ver_1, t_2)W(S_0)\ran\\
=\frac{\vesig_1\veL_1}{2m^2_1}\int_{-\infty}^\infty &dt_1&\int_{t_1}^\infty dt_2 (t_2-t_1) \varepsilon_{ijk}
\frac{r_k}{r^2}  \mbox{Tr}\lan g^2 H_i (\ver_1, t_1) E_j (\ver_1, t_2)W(S_0)\rangle
=-\int\Delta V_{SO}(r)dt_1.
\label{C20}
\end{eqnarray*}
Comparing the last formula with the definition
\be
\Delta V_{SO}(r)=\frac{\vesig_1\veL_1}{2m^2_1}\frac{1}{r}V'_1(r),
\ee
it is straightforward to extract the potential $V_1'(r)$. In particular, after expanding
the exponent in $W(S_0)$ and keeping only the lowest contribution, one arrives at
\be
\frac{dV_1^{EF}(r)}{dr}=-\int^\infty_0 tdt \int dw_4\int^r_0 d\lambda \lan E(2) E(w) H(1)\ran+O(\llan F^5\rran),
\label{C23}
\ee
where the surface element $ds_{n4}$ was written as
\be
ds_{n4}=dw_n dw_4=r_n d\beta dw_4=\frac{r_n}{r}d\lambda dw_4
\ee
and we defined the triple field correlator,
\be
\lan E(2)E(w)H(1)\ran\equiv\varepsilon_{ikl}f^{abc}g^3\lan E^a_i(\ver_1, t_2)E_k^b(\ver_1-\velam,w_4)H^c_l(\ver_1, t_1)\ran.
\label{C24}
\ee
Notice that each field operator is in fact sandwiched between the two parallel transporters (see Eq.~(\ref{partr})), like
\be
E(x)\equiv E(x,x_0)=\Phi(x_0, x)E(x)\Phi(x,x_0),
\label{D25}
\ee
which provide the overall gauge invariance of the field correlator. It is convenient to choose the reference point $x_0=(t_1,\ver_1)$.

We now employ the relation between the triple field correlator and the derivative of the bilocal field corelator \cite{SSCs}:
\be
\frac{\partial}{\partial x_n} D(x-y) = \frac12 \int^x_y dz_n\alpha (z) \langle E(x) E(y) H(z)\rangle,\quad
\alpha(z)=\frac{z-y}{x-y}.
\label{C28}
\ee
By choosing
$$
n=4,\quad t=z-y,\quad y=t,\quad z=t_2,\quad x=w_4,
$$
one can rewrite Eq.~(\ref{C23}) as
\begin{eqnarray}
\frac{dV_1^{EF}(r)}{dr}&=&-\int^\infty_0 tdt\int_{t_1}^\infty d w_4\int_0^r d\lambda\langle E(2)E(w)H(1)\rangle\nonumber\\
&=&-2\int_{t_1}^\infty (w_4-t_1)\frac{\partial}{\partial w_4}D^H(w_4-t_1,\lambda)dw_4 d\lambda\label{V1prEF}\\
&=&-2\int^\infty_{t_1}D(w_4-t_1, \lambda) dw_4 d\lambda=-2\int_0^\infty d\nu\int^r_0 d\lambda D^H(\nu,\lambda).\nonumber
\end{eqnarray}
This result coincides with the FCM expression given by Eq.~(\ref{C9}), if one neglects the term containing $\lambda/r$.
As seen from Eq.~(\ref{V1prEF}), this term does not appear in the lowest order of the Eichten--Feinberg representation.
It is easy to trace the root of this term. In view of the equality (\ref{C7}), one can write:
\be
\frac{dV_1^{EF}(r)}{dr}+O(\llan F^5\rran)=\frac{dV_1^{FC}(r)}{dr}+O(\llan F^4\rran)
\ee
or, keeping only the lowest correction,
\be
\frac{d}{dr}[V_1^{EF}(r)-V_1^{FC}(r)]=O(\llan F^4\rran).
\ee
In other words, the appearance of the correction $\lambda/r$ signifies contributions of higher-order terms in the cluster expansion.
This correction is clearly of order of $T_g/r$ (see also Eq.~(\ref{V1cr}) below) and thus it decreases with the decrease of the correlation length $T_g$. The
two representations, the Field Correlator representation and the Eichten--Feinberg one, coincide therefore in the string limit of QCD when $T_g\to 0$.

A similar consideration is valid for the potential $V_2'(r)$, so that one arrives at the following formulae for the two potentials in the
Eichten--Feinberg representation (to be compared with the FCM formulae given by Eq.~(\ref{Vseq})):
\beq
\frac{dV^{EF}_1(r)}{dr}&=&-2\int^\infty_{0} d\nu\int^r_0d\lambda D^H(\lambda,\nu),\label{300}\\
\frac{dV^{EF}_2(r)}{dr}&=&r\int^\infty_{0} d\nu D^H_1(r,\nu).
\label{310}
\eeq

\section{Spin-dependent potentials and field correlators on the lattice}\label{S2}

\subsection{Spin-dependent potentials in heavy quarkonia: The gluonic correlation length}

Recently spin-dependent potentials in a heavy quarkonium were
measured on the lattice in quenched approximation and without cooling \cite{Komas}. Several different lattice
configurations were used in Ref.~\cite{Komas}, and the accuracy of these
calculations is essentially better than in previous lattice
calculations \cite{lt1,Bali1}. In this paper we
compare the predictions of the FCM with the results from
Ref.~\cite{Komas} obtained for the lattice size $20^3 40$ and the
bare gauge coupling $\beta=6.0$ (this corresponds to the lattice
spacing $a=0.093$ fm). The fits to the data found in
Ref.~\cite{Komas} read:
\beq
V_{0\rm fit}'(r)&=&\sigma+\frac{c}{r^2},\quad\sigma a^2=0.0468(2),\quad c=0.297(1),\nonumber\\
V_{1\rm fit}'(r)&=&-\sigma_{\rm v1},\quad\sigma_{\rm v1}a^2=0.0362(4),\nonumber\\
V_{2\rm fit}'(r)&=&\sigma_{\rm v2}+\frac{c_{\rm v2}}{r^2},\quad\sigma_{\rm v2} a^2=0.0070(7),\quad c_{\rm v2}=0.288(7),\label{Kfits}\\
V_{3\rm fit}(r)&=&\frac{3c_{\rm v3}}{r^3},\quad c_{\rm v3}=0.214(2),\nonumber\\
V_{4\rm fit}(r)&=&-g'm_g^2\frac{e^{-m_g r}}{r}+4\frac{\sigma_{\rm v4}}{r},
\;m_g a=1.16(\mbox{fixed}),\; g'=0.292(12),\;\sigma_{\rm v4} a^2=0.015(3)\nonumber.
\eeq

\begin{table}[t]
\begin{ruledtabular}
\begin{tabular}{cccccc}
Potential&$V_0'(r)$&$V_1'(r)$&$V_2'(r)$&$V_3(r)$&$V_4(r)$\\
\hline
$\sigma$, GeV$^2$&0.22&0.17&0.03&$-$&0.07\\
\hline
$\alpha_S$&0.22&$-$&0.22&0.16&$-$
\end{tabular}
\caption{The set of parameters, in physical units, extracted from the fits (\ref{Kfits}) built for the lattice data in Ref.~\cite{Komas}.}\label{Kfts}
\end{ruledtabular}
\end{table}

\begin{table}[t]
\begin{ruledtabular}
\begin{tabular}{ccccc}
&$\alpha_S$&$\sigma$, GeV$^2$&$T_g$, fm&$T_g'$, fm\\
\hline
set 1&0.16&0.22&0.2&0.2\\
set 2&0.16&0.22&0.1&0.1\\
set 3&0.16&0.22&0.07&0.1\\
set 4&0.16&$-$&0&0\\
set 5&0.32&0.17&$-$&$-$\\
\end{tabular}
\caption{The sets of the FCM parameters for the spin-dependent potentials taken from Ref.~\cite{Komas}.
Eqs.~(\ref{Vseq}) and (\ref{V012eq}) are used for the sets~1-4 and set~5, respectively.}\label{ourfts}
\end{ruledtabular}
\end{table}

\begin{figure}[t]
\begin{center}
\epsfig{file=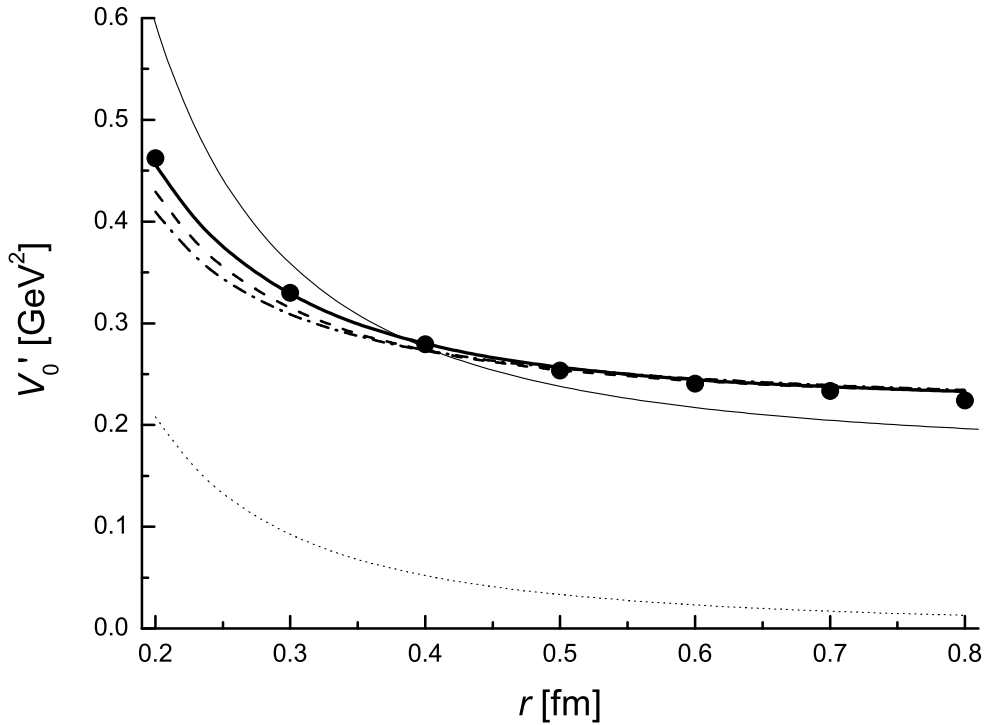,width=13cm}
\caption{The profile of the $V_0'(r)$ for the set~1 (dash-dotted line), set~2 (dashed line), set~3 (fat solid line), set~4 (doted line),
and set~5 (thin solid line). Lattice data are given by dots.}\label{V0fig}
\end{center}
\end{figure}

\begin{figure}[t]
\begin{center}
\epsfig{file=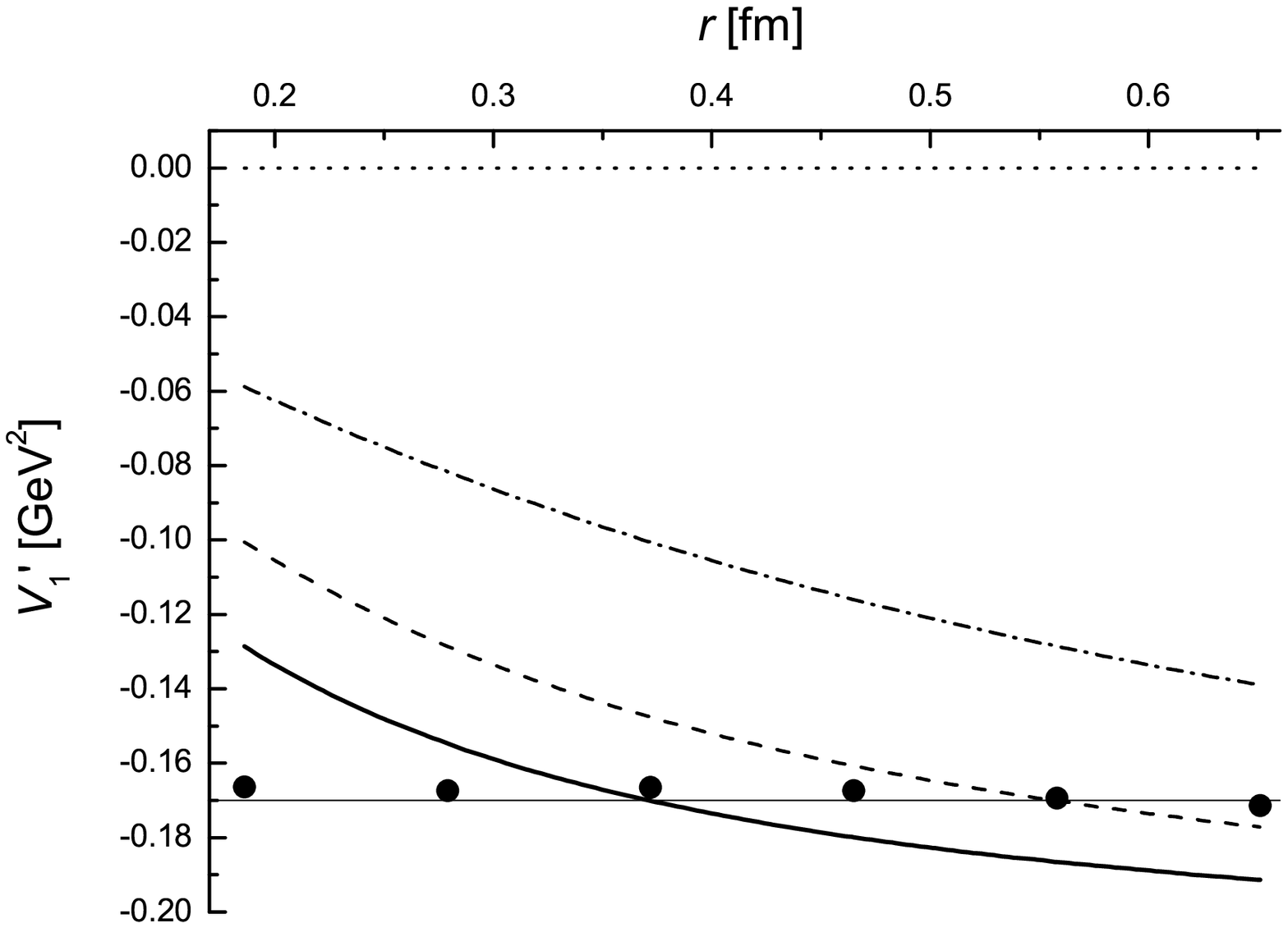,width=13cm}
\caption{The same as in Fig.~\ref{V0fig} but for $V_1'(r)$.}\label{V1fig}
\end{center}
\end{figure}

\begin{figure}[t]
\begin{center}
\epsfig{file=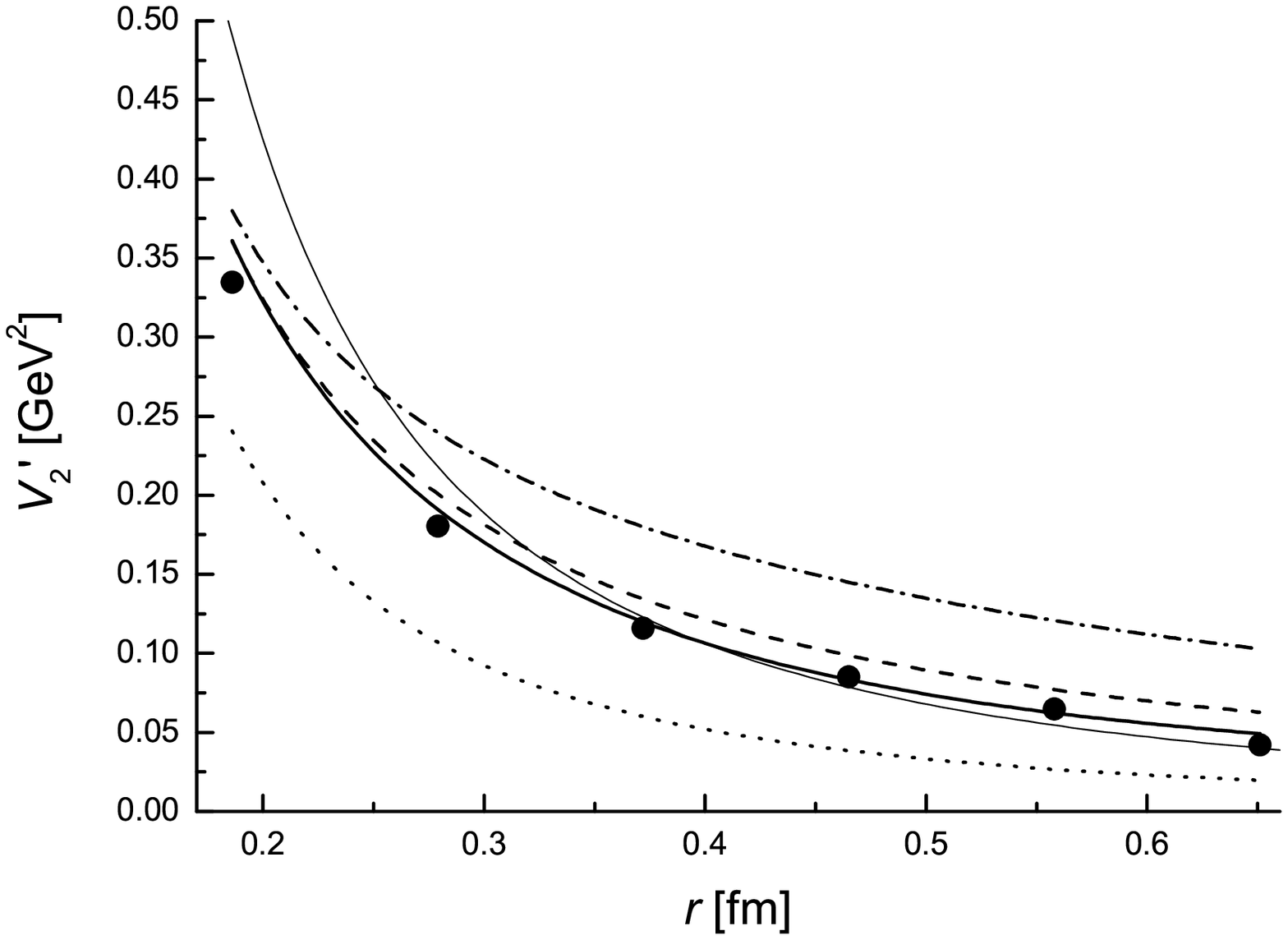,width=13cm}
\caption{The same as in Fig.~\ref{V0fig} but for $V_2'(r)$.}\label{V2fig}
\end{center}
\end{figure}

\begin{figure}[t]
\begin{center}
\epsfig{file=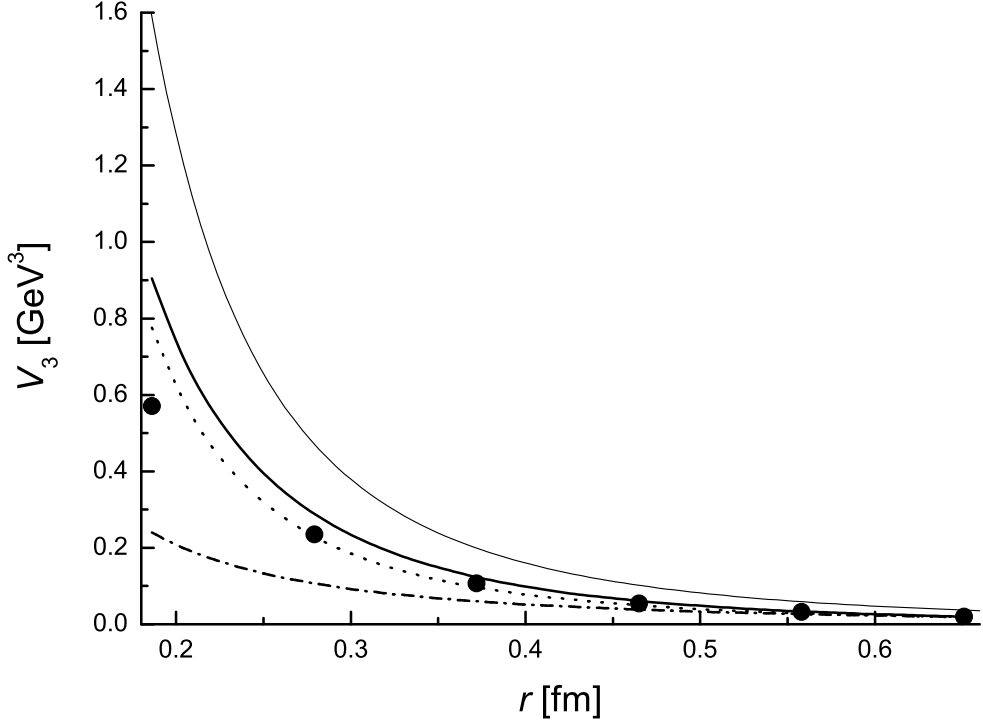,width=13cm}
\caption{The same as in Fig.~\ref{V0fig} but for $V_3(r)$. The curve for the set~2 coincides with that for the set~1.}\label{V3fig}
\end{center}

\end{figure}
\begin{figure}[t]
\begin{center}
\epsfig{file=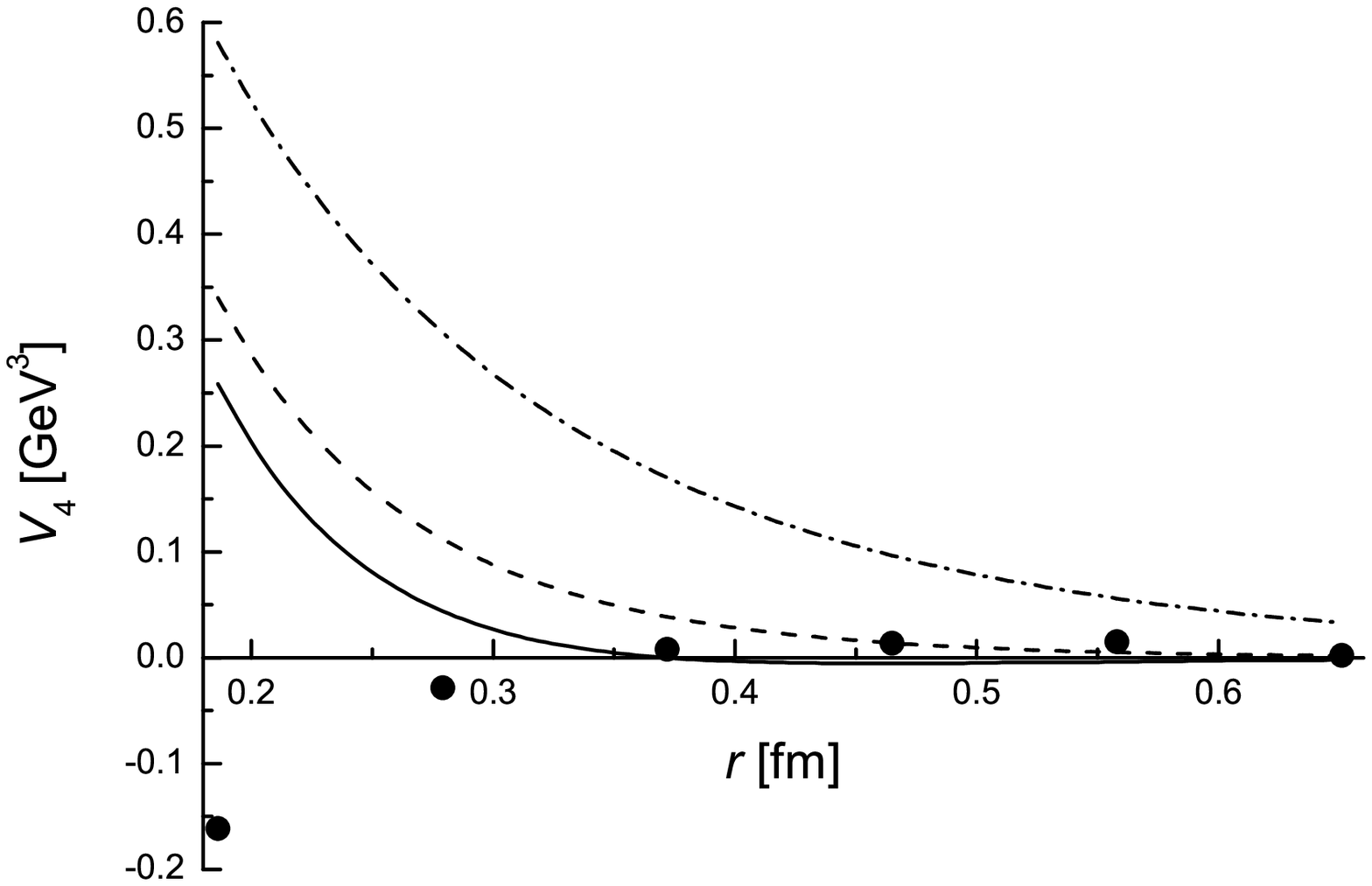,width=13cm}
\caption{The same as in Fig.~\ref{V0fig} but for $V_4(r)$. The sets~4 and 5 correspond to
delta-functions localised at $r=0$.}\label{V4fig}
\end{center}
\end{figure}

Parameters of these fits, in physical units, are listed in
Table~\ref{Kfts}. One can see that slightly different values of
the string tension and the strong coupling constant are used in
different fits in order to reproduce the data.

With the form of
the potentials derived in the FCM and given by analytic
expressions (\ref{Vseq}), (\ref{Dcor0}), and (\ref{D1cor0}) we are in a
position to compare the predictions of the FCM with the lattice data from Ref.~\cite{Komas}.
In the framework of the FCM, the spin-dependent potentials are defined with the help of
four parameters, $\{\alpha_S,\sigma,T_g,T_g'\}$, which possess clear physical meanings.
Notice that none of these four parameters can be treated as a fitting parameter, because their values are
strongly restricted by phenomenology.
Indeed, the string tension $\sigma$ defines, for example, the slope of the hadronic Regge trajectories and
can be extracted from the experimental data for the meson and baryon spectra. The behaviour of the strong coupling constant is dictated by QCD, while
the values of the correlation lengths are closely related to the gluelump spectrum. Indeed, the inverse of the correlation length
gives the mass of the lowest one-gluon ($M_0^{(1)}$) or two-gluon ($M_0^{(2)}$) gluelump \cite{simcor}:
\be
D_1^{NP}(u)\propto
\exp\left(-M_0^{(1)}|u|\right),\quad D(u)\propto
\exp\left(-M_0^{(2)}|u|\right), \quad
M_0^{(1)}=\frac{1}{T_g'},\quad M_0^{(2)}=\frac{1}{T_g}.
\ee
In the framework of the FCM these masses were calculated to be 2.5 GeV and 1.5 GeV, respectively \cite{simcor}
(for the value of the string tension $\sigma=0.18$ GeV$^2$). The relation between the gluonic correlation lengths and
the gluelump spectrum was also emphasized in Ref.~\cite{Nora123}. In particular, using the lattice data from Ref.~\cite{FM1}, the mass of the lowest
gluelump was calculated to be 0.9~GeV that corresponds to the gluonic correlation length 0.2~fm, which is consistent with the results from
Ref.~\cite{Tg}. As will be demonstrated below, the new lattice data \cite{Komas} give even smaller correlation length, so it would be interesting to
repeat the analysis of Ref.~\cite{Nora123} for the new data from Ref.~\cite{Komas}.

We, therefore, use the FCM form of the spin-dependent potentials, with the parameters varying in the range compatible with phenomenology,
as a motivated guess and compare our theoretical predictions with the lattice data.
We consider several such sets of the FCM parameters (as listed in Table~\ref{ourfts}), and our aim is to find the set which provides the best overall
description of the data.
For the sets~1-3 we use the full form of the
potentials (\ref{Vseq}); the set~4 demonstrates the relevance of
nonperturbative interactions since only perturbative part of the
potentials $V_n(r)$ ($n=0$-4) is retained in this case. Finally,
the data were approximated with the help of the asymptotic
large--distance potentials (\ref{V012eq}) (set~5). Comparison of the theoretical potentials with the lattice data is given by
Figs.~\ref{V0fig}-\ref{V4fig}. In addition, below we also compare the lattice data for the field
correlators with the predictions of the FCM.

One can deduce several conclusions from
Figs.~\ref{V0fig}-\ref{V4fig}. First of all, by comparing the
``purely perturbative" set~4 with the rest of the sets, one can
see that the data clearly indicates the presence of
nonperturbative contributions to the potentials, even at small
interquark separations, where the perturbative physics dominates.
Second, by comparing sets~1-3 with one another, one can conclude
that the data prefer small correlation lengths $T_g$ and $T_g'$: a
good description is achieved with them both being $\lesssim$ 0.1~fm.
Then, from the best set~3, the gluelump masses can be extracted as
$M_0^{(2)}=3$ GeV and $M_0^{(1)}=2$ GeV, which should be
confronted with the values 2.8 GeV and 1.7 GeV, respectively, obtained from the predictions of
Ref.~\cite{simcor} after the proper rescaling from $\sigma=0.18$
GeV$^2$ used in Ref.~\cite{simcor} to $\sigma=0.22$ GeV$^2$ taken
from our set~3.
For such small correlation lengths the ``large-distance" regime of
$r\gg T_g,T_g'$ happens already for quite small interquark
separations, which means that the asymptotic large-distance form
of the potentials (\ref{Vseq}) should give a good description of
the data. Indeed, our set~5, based on Eq.~(\ref{V012eq}),
approximates the data rather well (with the exception of the
potential $V_4(r)$). For obvious reasons, it fails at small
distances. In general, the sets~1-3 still give a better
description of the data, so one can conclude that the present data
allow one to study the ``anatomy" of the field correlators.

Let us now comment on the Gromes relation (\ref{Ge}). It was checked numerically in Ref.~\cite{Komas}
and found to be slightly violated. Indeed, using the data presented in
Eq.~(\ref{Kfits}) and in Table~\ref{Kfts}, one can find:
\be
V_0'(r)+V_1'(r)-V_2'(r)=[\sigma-\sigma_{\rm v1}-\sigma_{\rm v2}]+\frac{c-c_{\rm v2}}{r^2}=
0.02\left[1+\frac{0.018}{(r\;\mbox{[fm]})^2}\right]\;\mbox{GeV$^2$}\neq 0.
\ee
The source of this violation is not clear at the moment. It may stem from an inconsistent extraction on the lattice of the potentials entering the
Gromes relation which, by itself,
might be a purely lattice artifact to disappear in the continuum limit \cite{Komapr}. Indeed,
the Gromes relation is exact in the framework of the FCM ---
see the theoretical expressions for $V_n(r)$ $(n=0,1,2)$ given by our Eqs.~(\ref{Vseq}). The Eichten--Feinberg formulae also satisfy this
relation --- see Eqs.~(\ref{300}) and (\ref{310}) above for the form of the potentials $V_1'(r)$ and $V_2'(r)$; $V_0'(r)$ coincides with the FCM one.

Then, if one adopts the FCM definition (\ref{Vseq}) of the potentials $V_1'(r)$ and $V_2'(r)$,
it is easily seen from Fig.~\ref{V1fig}, that this violation is
mostly due to the constant behaviour of the lattice potential
$V_1'(r)$, which is impossible to get in the FCM formalism with
$T_g\gtrsim 0.05$ fm. Indeed, from the FCM form of the potential $V_1(r)$ given in Eq.~(\ref{Vseq}), one finds at $r\gg T_g$:
\be
V_1'(r)=-2\int_0^\infty d\nu\int_0^r d\lambda \left(1-\frac{\lambda}{r}\right)D(\lambda,\nu)
\approx -\sigma+\frac{2T_g^3}{r}\int_0^\infty d\xi \xi^2 D(\xi)=-\sigma+\frac{4\sigma T_g}{\pi r},
\label{V1cr}
\ee
where Eq.~(\ref{Dcor0}) was used for the profile of the correlator $D$ and only the leading correction in the expansion with the small parameter
$T_g/r$ was retained, which stems from the term $\lambda/r$ under the
integral. This correction is responsible for the weak $r$-dependence of the FCM potential $V_1'(r)$ depicted in Fig.~\ref{V1fig}.
Notice that a nonconstant behaviour of the potential $V_1'(r)$ was measured on the lattice in Ref.~\cite{Bali1}, though a very large uncertainty
in the data does not allow one to make definite conclusions on the actual $r$-dependence of this potential.

Consider now the Eichten--Feinberg definitions of the spin-dependent potentials \cite{EF} given by Eqs.~(\ref{C1})-(\ref{C3}) and which
are used in lattice calculations \cite{lt1,Bali1,Komas}.
The formula (\ref{C1}) implies that corrections, similar to that given in Eq.~(\ref{V1cr}), are not
included into the Eichten--Feinberg definition of the potential $[V^{\rm EF}_1(r)]'$, which is therefore a constant.
Exactly this constant behaviour is confirmed by the lattice calculations \cite{Komas} (see the fit $V_{1\rm fit}'(r)$ in the set (\ref{Kfits})) and,
according to Eq.~(\ref{300}),
the potential $V_1'(r)$ is purely nonperturative. On the contrary, from Eq.~(\ref{310}), one can see that, in the Eichten--Feinberg representation,
$V_2'(r)$ is of a purely perturbative nature. In the meantime,
a small, though nonvanishing nonperturbative contribution to the potential $V_2(r)$ was found on the lattice ---
a nonzero value of the $\sigma_{\rm v2}$ in the $V_{2\rm fit}'(r)$.
Thus we conclude that the problem with the Gromes relation on the lattice deserves further investigation with a special attention payed to the surface used in the Wilson
loop and to the contribution of higher correlators, since the root of the discrepancy may reside exactly here, as was explained before.

Finally, there is a certain contradiction between the theoretical predictions and the lattice data for the potential $V_4(r)$. Our form of
the potential $V_4(r)$ given by Eq.~(\ref{Vseq}) is consistent with the delta-functional form of this potential in the limit of the vanishing
correlation length (see Eq.~(\ref{V012eq})). This should not come as a surprise since the perturbative
part of the correlator $D_1(u)$ reproduces the OGE.
Although, for finite values of the correlation length, $V_4(r)$ is smeared and can become negative,
the small-$r$ behaviour of our theoretical curves  ($r<0.3 $ fm) is different
from that of the lattice data, so this question also deserves additional investigation.

\subsection{Gaussian field correlators on the lattice}\label{latcor}

\begin{figure}[t]
\begin{center}
\epsfig{file=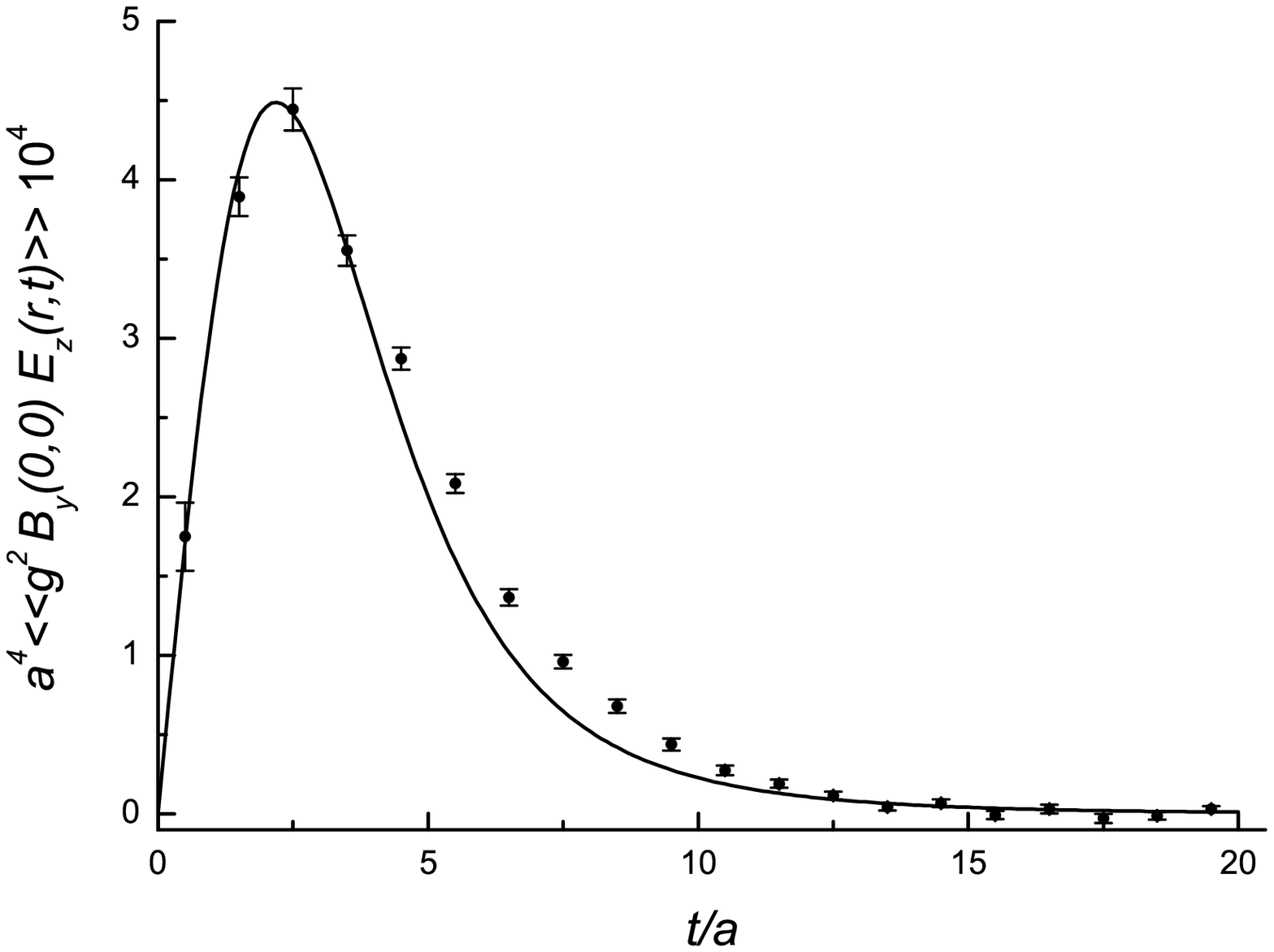,width=8cm}\epsfig{file=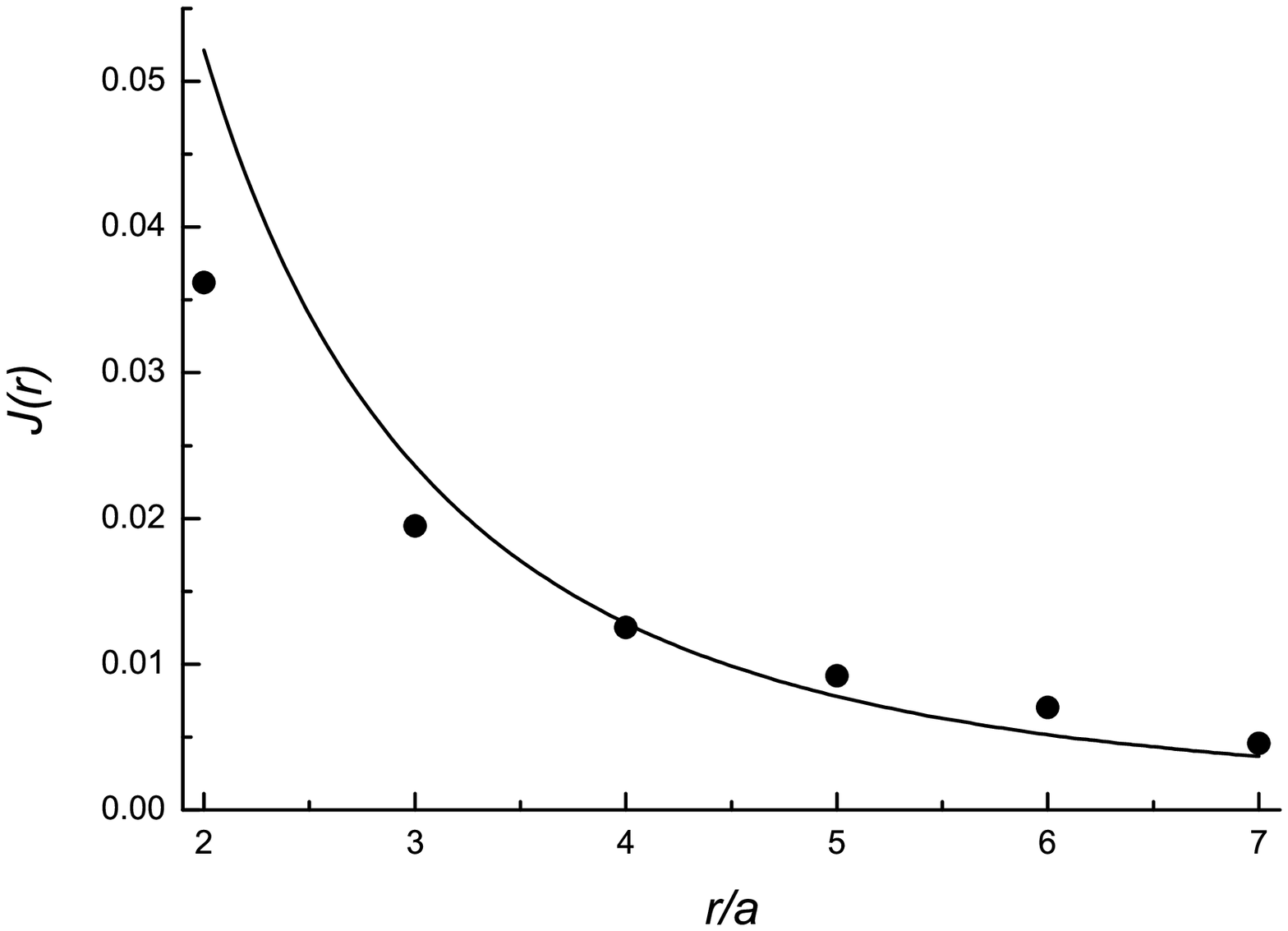,width=8cm}
\caption{The field correlator $a^4\llan g^2 B_y(0,0)E_z(r,t)\rran$ at $r/a=5$ (first plot) and the integral
$J(r)=a^2\int_0^\infty\llan g^2 B_y(0,0)E_z(r,t)\rran tdt$ (second plot). Lattice data are given by dots.}\label{ByEzfig}
\end{center}
\end{figure}

\begin{figure}[t]
\begin{center}
\epsfig{file=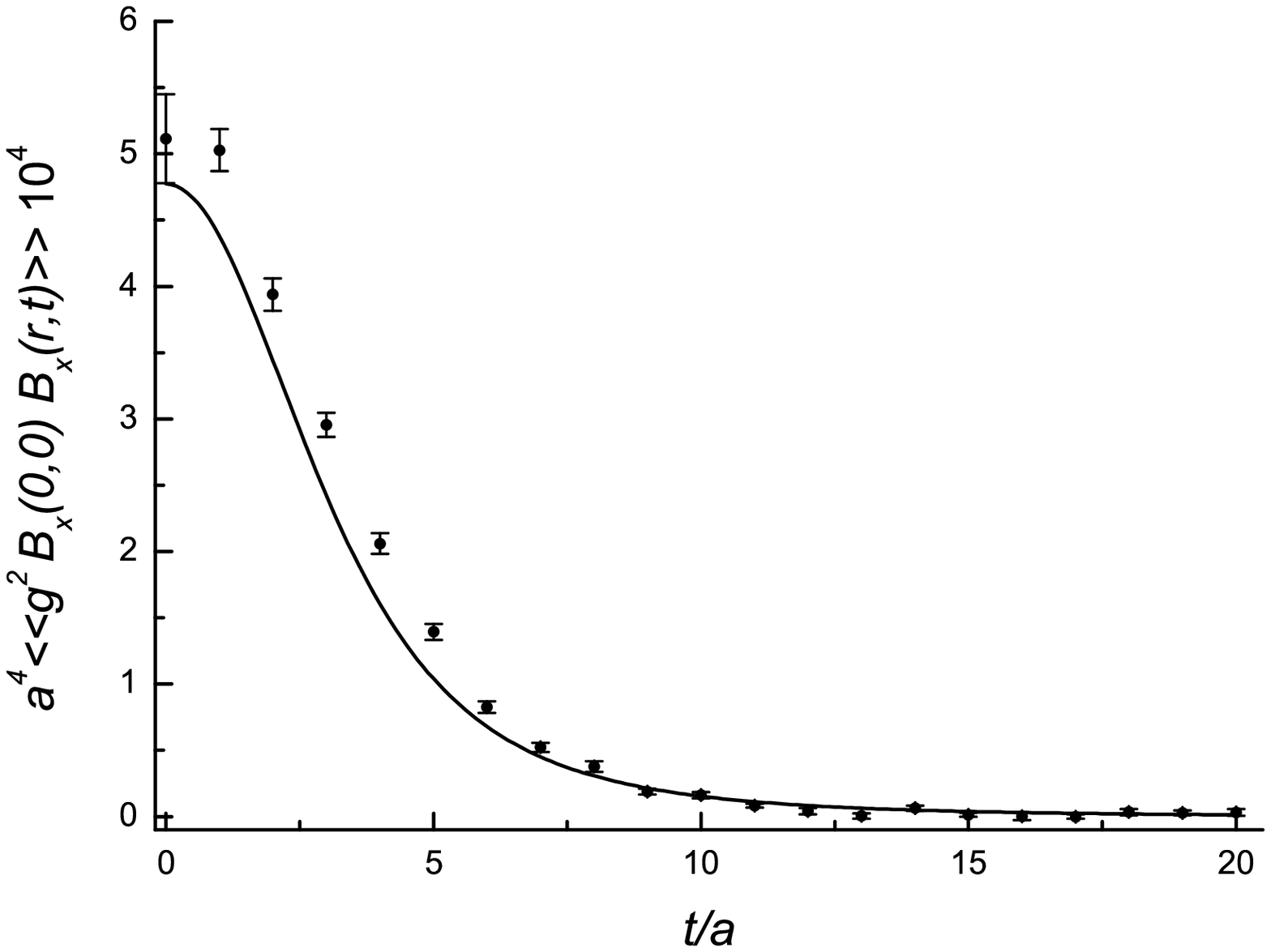,width=8cm}\epsfig{file=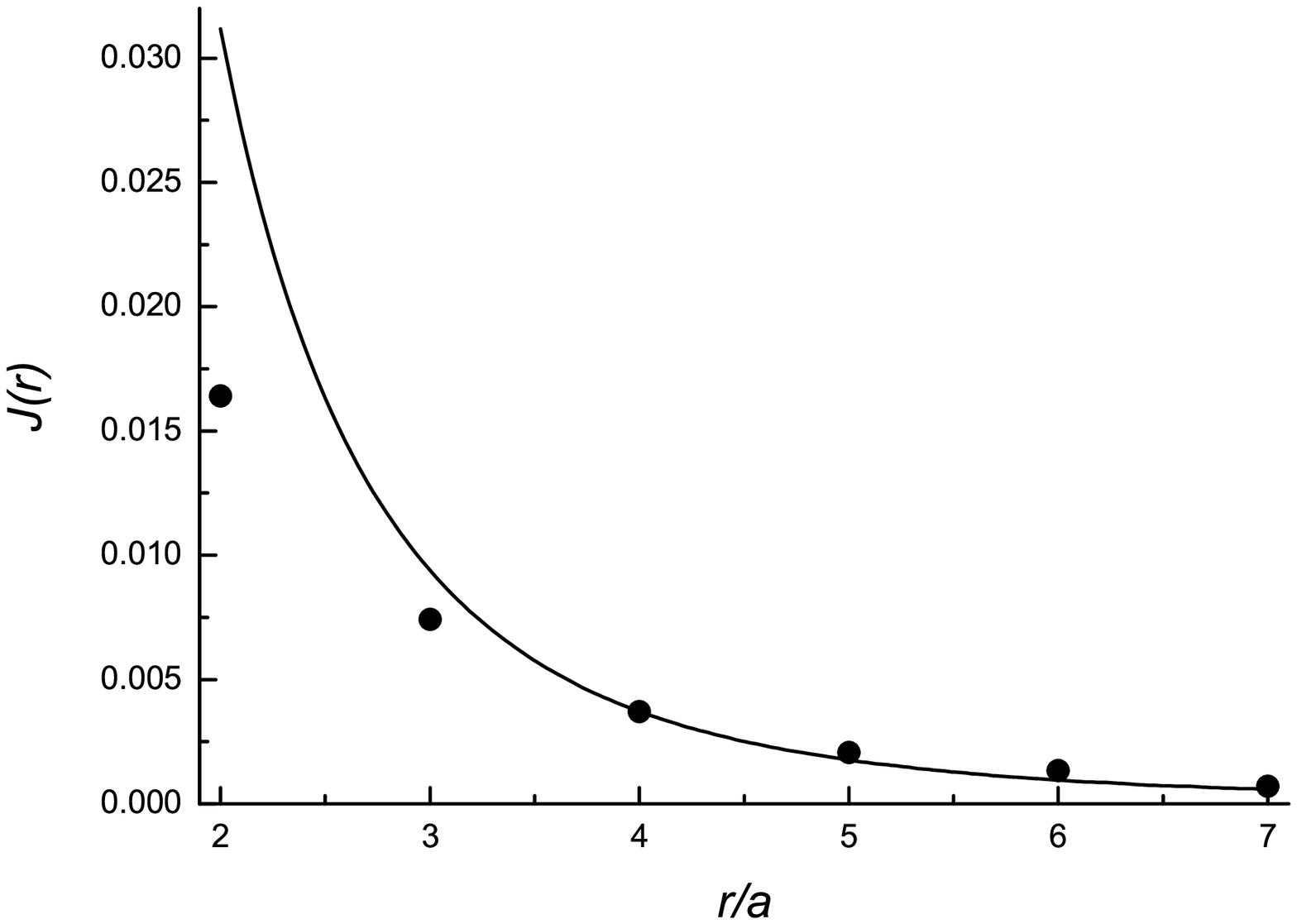,width=8cm}
\caption{The field correlator $a^4\llan g^2 B_x(0,0)B_x(r,t)\rran$ at $r/a=5$ (first plot) and the integral
$J(r)=a^3\int_0^\infty\llan g^2 B_x(0,0)B_x(r,t)\rran dt$ (second plot). Lattice data are given by dots.}\label{BxBxfig}
\end{center}
\end{figure}

\begin{figure}[t]
\begin{center}
\epsfig{file=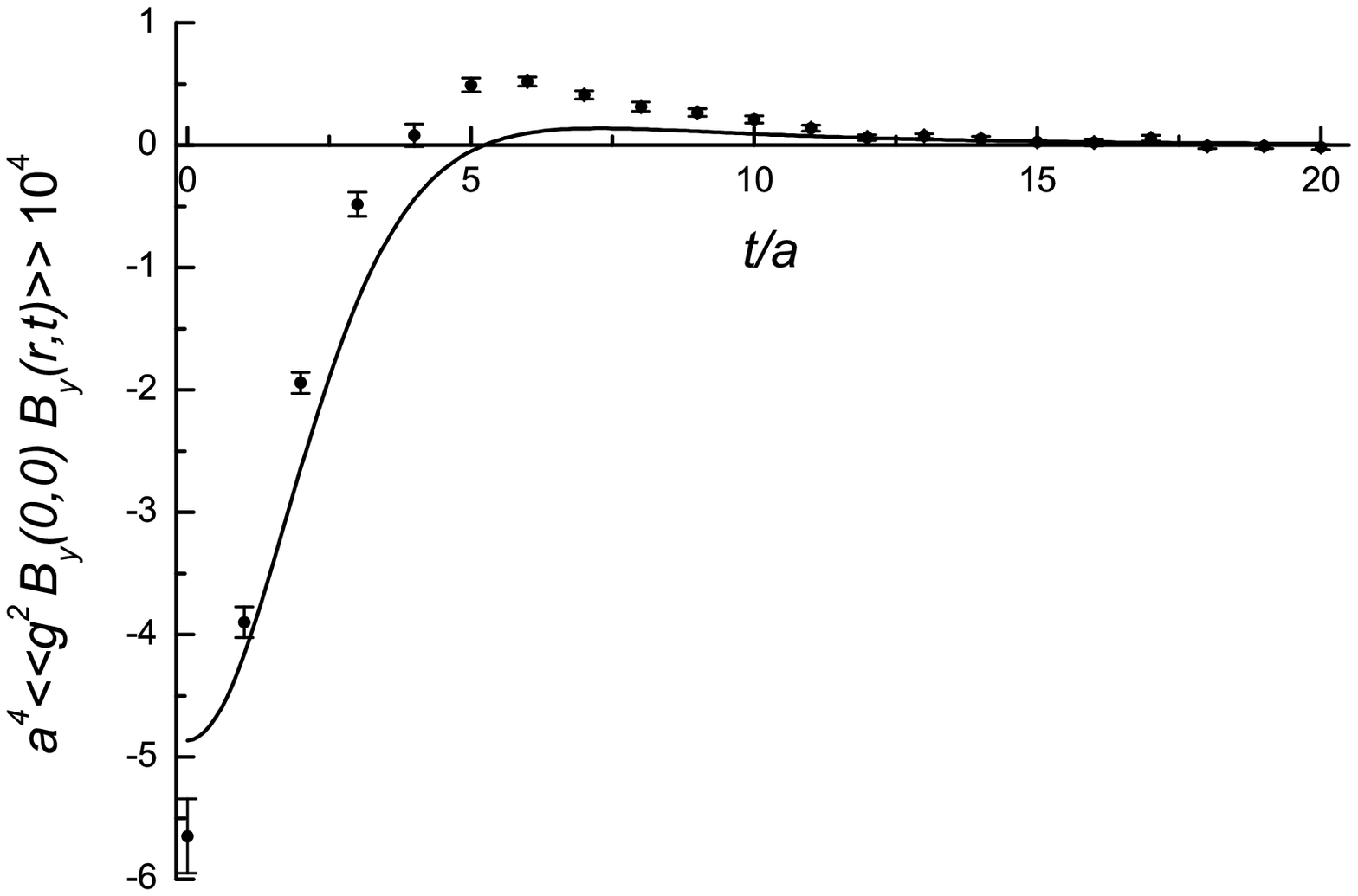,width=8cm}\epsfig{file=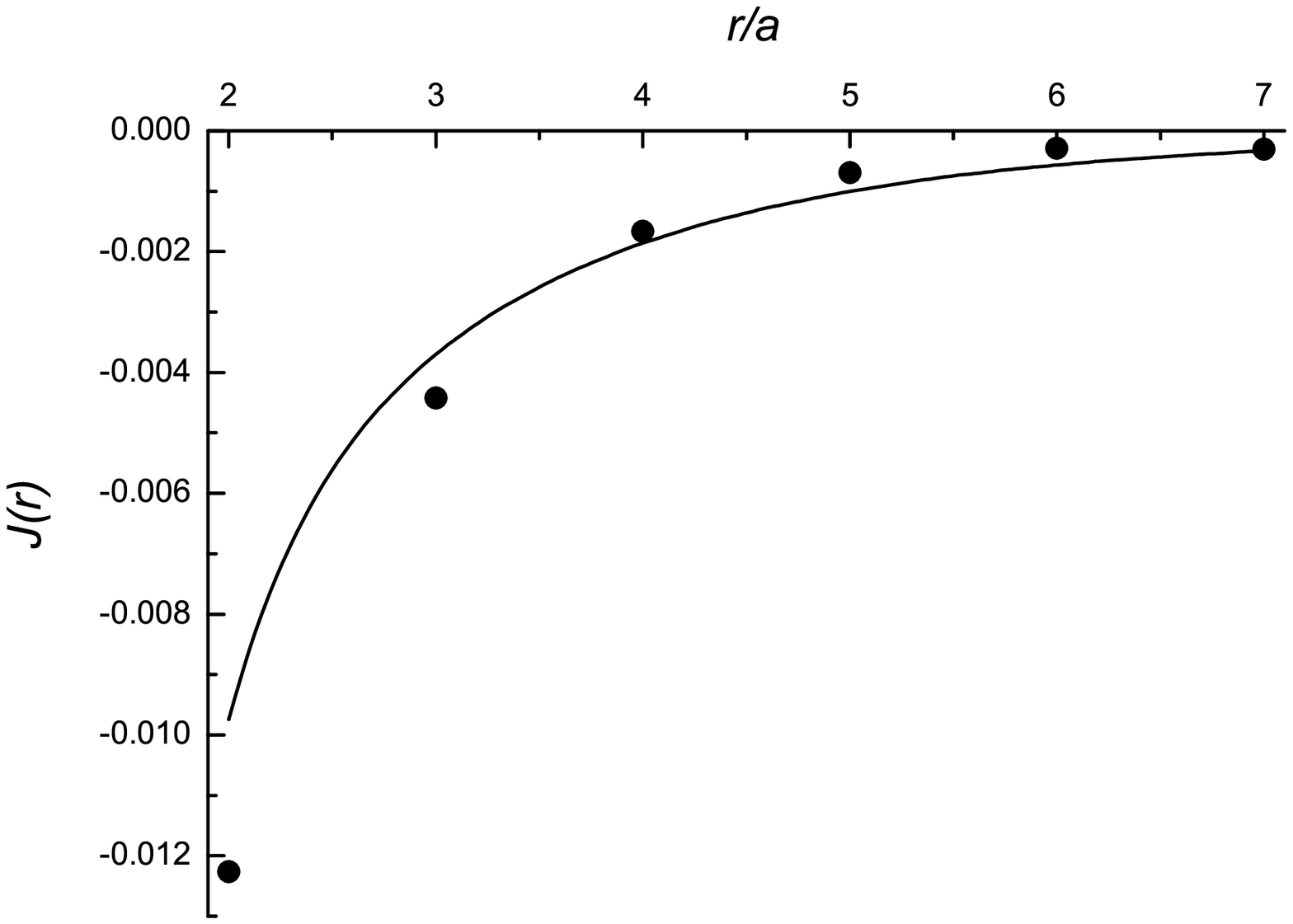,width=8cm}
\caption{The field correlator $a^4\llan g^2 B_y(0,0)B_y(r,t)\rran$ at $r/a=5$ (first plot) and the integral
$J(r)=a^3\int_0^\infty\llan g^2 B_y(0,0)B_y(r,t)\rran dt$ (second plot). Lattice data are given by dots.}\label{ByByfig}
\end{center}
\end{figure}

In this chapter we compare the form of the Gaussian field correlators given by the FCM with the lattice data taken from Ref.~\cite{Komas}.
As before, we consider the data for the lattice $20^340$ with $\beta=6.0$.
In particular, we evaluate the correlators of the two plaquettes inserted at the positions
$(0,0,0,0)\equiv (0,0)$ and $(r,0,0,t)\equiv(r,t)$.
Then, with the help of Eqs.~(\ref{Hs0}) and (\ref{HE0}), it is easy to find the correlators measured on the lattice
in the form ($u^2=r^2+t^2$):
\beq
\frac{1}{N_C}\llan g^2 B_y(0,0)E_z(r,t)\rran&=&-rt\frac{\partial D_1(u)}{\partial r^2},\label{BE1}\\
\frac{1}{N_C}\llan g^2 B_x(0,0)B_x(r,t)\rran&=&D(u)+D_1(u),\label{BB1}\\
\frac{1}{N_C}\llan g^2 B_y(0,0)B_y(r,t)\rran&=&D(u)+D_1(u)+r^2\frac{\partial D_1(u)}{\partial r^2},\label{BB2}
\eeq
where, in order to comply with the definitions and notations used in Ref.~\cite{Komas}, we reverted the sign of the correlator
$\llan g^2 B_yE_z\rran$ (to be compared with Eq.~(\ref{HE0})) and denoted the magnetic field as $\veB$.
Notice that the lattice field strength correlator is calculated as
a matrix element of an operator defined in (potential) nonrelativistic QCD and simulated on the lattice, which is to be multiplied by
the appropriate renormalization factor in order to give the corresponding correlator in continuum. Although there is an estimate for the magnetic
factor $Z_B$ \cite{alpha}, the electric factor $Z_E$ is unknown yet. Since such renormalisation factors do not change the shape of the
correlators, we adopt the following strategy. The profile functions $D(u)$ and $D_1(u)$ are taken in the form of Eqs.~(\ref{Dcor0}) and (\ref{D1cor0}),
with the set of parameters providing the best overall description of the spin-dependent potentials (the set~3 --- see Table~\ref{ourfts} and
Figs.~\ref{V0fig}-\ref{V4fig}). Then the right-hand sides of Eqs.~(\ref{BE1})-(\ref{BB2}) are maltiplied by the normalisation factors ${\cal N}_1$,
${\cal N}_2$, and ${\cal N}_3$, respectively. Notice however that these three factors are not independent. Indeed, clearly,
${\cal N}_2={\cal N}_3$ since only magnetic fields are involved in both correlators. On the other hand,
from Eqs.~(\ref{BE1})-(\ref{BB2}) one can derive easily the relation between the correlators which is independent of any particular
choice of the profile functions and fitting constants, namely:
\be
t[\llan B_x(0,0)B_x(r,t)\rran-\llan B_y(0,0)B_y(r,t)\rran]=r\llan B_y(0,0)E_z(r,t)\rran.
\label{rel0}
\ee
This equation establishes the relation between the factors ${\cal N}_1$ and ${\cal N}_2$. In other words, only one normalisation factor of the three
is independent, for example, ${\cal N}_1$. We use it as a free parameter.

The correlators (\ref{BE1})-(\ref{BB2}) (multiplied by the fitted values of the normalisation factors ${\cal N}_1\approx 1.1$,
${\cal N}_2={\cal N}_3\approx 0.6$) together with the lattice data are plotted in
Figs.~\ref{ByEzfig}-\ref{ByByfig}. From these figures one can see a reasonable agreement between the predictions of the FCM and the lattice data.
The field renormalisation factors can be calculated then as
\be
Z_B=\frac{1}{\sqrt{{\cal N}_2}}\approx 1.3,\quad Z_E=\frac{\sqrt{{\cal N}_2}}{{\cal N}_1}\approx 0.7.
\label{factors}
\ee

Two comments are in order here. The first comment concerns the small-$r$ behaviour of the integrated correlators $J(r)$
(see Figs.~\ref{ByEzfig}-\ref{ByByfig}), the agreement with the data
worsening as $r\to 0$. This may be related to two facts. On one hand, the contribution of higher correlators at small $r$'s is suppressed by extra
powers of $r$, which might translate into extra powers of the ratio $r/a$ on the lattice. Therefore, one should suspect
discretisation errors of lattice to become significant at short distances. On the other hand, at small interquark separations,
the perturbative physics dominates, with the behaviour of
the correlators being essentially dictated by the running strong coupling constant $\alpha_s(r)$. In our approach adopted in this paper
we neglect running of the $\alpha_s$
which appears to be a good approximation at large and moderate distances and fails only at $r\lesssim 0.2$ fm.
An accurate description of the data at such small distances would require not only considering a running coupling constant but also an appropriate
modification of the profile functions of the correlators in order to warrant a regular behaviour of the latter at the origin, as was discussed before.
For most applications, however, the large-distance behaviour of the correlators is essential. In particular,
in the present paper, we apply the FCM to the spin-dependent potentials, where also large distances are important.

The second comment is about the renormalisation factors $Z_E$ and $Z_B$ estimated above and computed on the lattice in Ref.~\cite{Komas}. Although our
colour-magnetic factor $Z_B\approx 1.3$ appears in a satisfactory agreement with the lattice result $Z_B^{\rm lat}\approx 1.6\div 1.7$,
the corresponding colour-electric
factor $Z_E^{\rm lat}\approx 1.5\div 1.6$ differs substantially from the result given in Eq.~(\ref{factors}). Notice that one should not
require exact equality of unobservable quantities, such as the field renormalisation factors, measured on the lattice and calculated in the
continuum limit, because of the discretisation errors and finite volume effects on the lattice. Besides that quite different techniques are used in these two
approaches, so that establishing a one-to-one correspondence between auxiliary quantities may not be possible.
Meanwhile, regardless of the
explicit values taken by the aforementioned renormalisation factors, their ratio is fixed by the relation (\ref{rel0}).
This relation is model-independent since it follows from the most general parametrisation of the Gaussian field correlator (\ref{Dcordef})
compatible with its Lorentz structure. Then, for the ``bare" field correlators calculated in Ref.~\cite{Komas} to satisfy this relation,
the ratio has to be $Z_B/Z_E\approx 2$ (see Eq.~(\ref{factors})), while it was measured on the lattice to be close to unity \cite{Komas}. This remains
an open question which requires further investigation. Concluding this discussion let us mention that,
since the renormalisation factors for the fields are expected to be calculated directly from QCD,
then the relation (\ref{rel0}) and similar relations between field correlators
can be used as a cross-check of the calculated factors or, until these calculations are performed, to relate
such renormalisation factors with one another.

\section{Quark masses and the effects of the string moment of inertia}\label{muvsm}

It was mentioned before that the correct definition of the quark masses is of great importance for the correct treatment of the spin-dependent potentials
in quarkonia. Our purpose here is to define the formal quantities denoted before as $\bar{m}_i$ ($i=1,2$). This is especially important for light
quarks since the very expansion in the inverse powers of the pole quark mass is meaningless in this case. In addition we shall take into account
the string contribution to the canonical angular momentum.

We return now to the quark--antiquark Greens's function (\ref{6s}) and consider the spin-independent part of the meson action.
The kinetic terms for the quarks are given by Eq.~(\ref{7s}), whereas the spin-independent interaction comes from the Wilson loop (\ref{Smin}).

Using the straight-line string ansatz and considering the system
in the laboratory reference frame, we synchronise the quark times,
\be
x_{10}=x_{20}=t,
\ee
and thus arrive at the standard
Lagrangian of the QCD string with quarks at the ends \cite{DKS}:

\be
L=-m_1\sqrt{1-\dot{\vex}_1^2}-m_2\sqrt{1-\dot{\vex}_2^2}-\sigma
r\int_0^1d\beta\sqrt{1-[\ven\times(\beta\dot{\vex}_1+(1-\beta)\dot{\vex}_2)]^2},
\label{L2}
\ee
where $\ver=\vex_1-\vex_2$, $\ven=\ver/r$. The
square roots in the Lagrangian (\ref{L2}) can be conveniently
treated with the help of the einbein (auxiliary) field formalism
\cite{ein}. In this formalism, an extra degree of freedom --- the
einbein field --- is introduced,
\be
m\sqrt{1-\dot{\vex}^2}\to\frac{m^2}{2\mu}+\frac{\mu}{2}-\frac{\mu\dot{\vex}^2}{2},
\ee
where the original form of the Lagrangian is restored as soon
as the extremum in the einbein $\mu$ is taken. Thus one reduces
the relativistic kinematics to the nonrelativistic one. However, the
treatment of the system with the help of the einbein field
formalism remains a full relativistic treatment since the entire
set of relativistic corrections is effectively performed by taking
extremum in the einbein field. The problem of the centre-of-mass
motion separation becomes a trivial problem with the einbeins
introduced (similarly to the einbeins $\mu_i$ ($i=1,2$) for the
quarks, a continuous einbein $\nu(\beta)$ is introduced in order
to simplify the string term in Eq.~(\ref{L2})). The result reads
(for the sake of brevity we omit the part of the Lagrangian
responsible for the centre-of-mass motion) \cite{DKS,GD}:
\be
L=-\frac{m^2_1}{2\mu_1}-\frac{m^2_2}{2\mu_2}-\frac{\mu_1}{2}-\frac{\mu_2}{2}-
\int^1_0\frac{\nu}{2}d\beta-\int^1_0\frac{\sigma^2r^2}{2\nu}d\beta
+\frac12\mu(\dot{\ver}\ven)^2+\frac12\tilde{\mu}[\dot{\ver}\times\ven ]^2,
\label{L13}
\ee
where
\be
\veR=\zeta_1\vex_1+(1-\zeta_1)\vex_2,\quad
\ver=\vex_1-\vex_2,\quad\zeta_1=\frac{\mu_1+\int_0^1\beta\nu d\beta}{\mu_1+\mu_2+ \int^1_0 \nu d\beta}, \quad\zeta_2=1-\zeta_1,
\label{Rr}
\ee
and we have defined the reduced masses for the angular and radial motion separately:
\be
\mu=\frac{\mu_1\mu_2}{\mu_1+\mu_2},
\label{murad}
\ee
\be
\tilde{\mu}=\mu_1(1-\zeta_1)^2+\mu_2\zeta_1^2+\int^1_0(\beta-\zeta_1)^2\nu d\beta.
\label{muang}
\ee
The physical meaning of the variables $\mu_i$ ($i=1,2$) is the average kinetic energy of the $i$-th
particle in the given state, namely, $\mu_i=\lan\sqrt{\vep^2+m^2_i}\ran$ (see the discussion in
Refs.~\cite{DKS,KNS}). The continuous einbein variable $\nu(\beta)$ has the meaning of the QCD string energy density
\cite{DKS}.

Following the standard procedure, we build now the canonical momentum as
\be
\vep=\frac{\partial L}{\partial \dot{\ver}}=\mu(\ven\dot\ver)\ven+\tilde{\mu}(\dot\ver-\ven(\ven\dot{\ver})),
\label{v16}
\ee
with its radial component and transverse component being
\be
(\ven\vep)=\mu(\ven\dot{\ver}),\quad [\ven\times\vep]=\tilde{\mu}[\ven\times\dot{\ver}],
\label{160}
\ee
respectively. Thus we arrive at the spin-independent part of the Hamiltonian \cite{DKS}:
\be
H=\sum_{i=1}^2\left[\frac{m_i^2}{2\mu_i}+\frac{\mu_i}{2}\right]+\int^1_0d\beta\left[\frac{\sigma^2r^2}{2\nu}+\frac{\nu}{2}\right]
+\frac{p_r^2}{2\mu}+\frac{\veL^2}{2\tilde{\mu}r^2}.
\label{Hm}
\ee

One can see therefore that the angular-momentum-dependent term in
the Hamiltonian (\ref{Hm}) contains the total moment of inertia
$\tilde{\mu}$ which includes both the effective masses of the
quarks and also the proper string moment of inertia. Notice that
the difference between $\mu$ and $\tilde{\mu}$ in the last term in
Eq.~(\ref{Hm}) gives rise to the so-called string correction in
the spin-independent Hamiltonian \cite{Sim2,KNS} and, finally, to
the correct Regge slope $M^2=2\pi\sigma J$ \cite{MNS}. Clearly,
the mass $\tilde{\mu}$ appears every time the angular
momentum operator is involved.

We now turn back to the spin-dependent potentials, namely, to the spin--orbital interaction which must be affected by the string inertia, as was
explained before. With the help of Eqs.~(\ref{Rr}) and (\ref{160}) one can find that the quantity ${\bm\rho}$,
which enters $ds_{ik}$ and which gives rise to the angular-momentum-dependent interquark interaction --- see Eq.~(\ref{19a}), takes the form:
\be
{\bm \rho}=[\ver\times(\beta\dot{\vex}_1+(1-\beta)\dot{\vex}_2)]=(\beta-\zeta_1)[\ver\times\dot{\ver}]=
(\beta-\zeta_1)\frac\veL{\tilde{\mu}}.
\label{a250}
\ee

It is straightforward now to derive the spin-dependent quark--antiquark interactions in the full form similar to that given in Eq.~(\ref{SO0}).
It is important to stress however that, unlike Eq.~(\ref{SO0}), this result is not due to the $1/m$ expansion but is obtained with the only
approximation made being the Gaussian approximation for the field correlators.

Finally, for the spin-dependent interactions, the following modification of Eq.~(\ref{SO0}) can be obtained (see
Appendices~\ref{colel} and \ref{colmagn} for the details):
\beq
\left(\frac{{\bm\sigma}_1\veL}{4{\bar m}_1^2}+\frac{{\bm\sigma}_2\veL}{4{\bar m}_2^2}\right)\frac1r\frac{dV_0}{dr}\to
\frac{1}{2r}\int_0^\infty d\nu\int_0^r d\lambda \left[\vphantom{\frac12}D+D_1+(\lambda^2+\nu^2)\frac{\partial D_1}{\partial\nu^2}\right]\hspace*{3cm}\nonumber\\
\times\left[(1-\zeta_1)\frac{{\bm\sigma}_1\veL}{\mu_1\tilde{\mu}}+(1-\zeta_2)\frac{{\bm\sigma}_2\veL}{\mu_2\tilde{\mu}}\right]\nonumber,\\[2mm]
\left(\frac{{\bm\sigma}_1\veL}{4{\bar m}_1^2}+\frac{{\bm\sigma}_2\veL}{4{\bar m}_2^2}\right)\frac2r\frac{dV_1}{dr}\to-\frac1r\int^\infty_0d\nu\int^r_0d\lambda D
\left(1-\frac{\lambda}{r}\right)\left[(1-\zeta_1)\frac{{\bm\sigma}_1\veL}{\mu_1\tilde{\mu}}+(1-\zeta_2)\frac{{\bm\sigma}_2\veL}{\mu_2\tilde{\mu}}\right],
\label{V02}\\[2mm]
\frac{({\bm\sigma}_1+{\bm\sigma}_2)\veL}{2{\bar m}_1{\bar m}_2}\frac1r\frac{dV_2}{dr}\to
\frac{1}{r^2}\int^\infty_0d\nu\int^r_0\lambda d\lambda \left[\vphantom{\frac12}D+D_1
+\lambda^2\frac{\partial D_1}{\partial\lambda^2}\right]\frac{(\vesig_1+\vesig_2)\veL}{\tilde{\mu}}\left(\frac{\zeta_1}{\mu_1}\right),\nonumber\\
\frac{(3(\vesig_1\ven)(\vesig_2\ven)-\vesig_1\vesig_2)}{12\bar{m}_1\bar{m}_2}V_3(r)\to
-2r^2\frac{\partial}{\partial r^2}\int_0^\infty d\nu D_1(r,\nu)\frac{(3(\vesig_1\ven)(\vesig_2\ven)-\vesig_1\vesig_2)}{12\mu_1\mu_2}\nonumber\\
\frac{\vesig_1\vesig_2}{12\bar{m}_1\bar{m}_2}V_4(r)\to6\int_0^\infty d\nu\left[D(r,\nu)+\left[1+\frac23r^2\frac{\partial}{\partial\nu^2}\right]D_1(r,\nu)\right]
\frac{\vesig_1\vesig_2}{12\mu_1\mu_2},\nonumber
\eeq
where the masses $\bar{m}_i$ are replaced by $\mu_i$ and $\tilde\mu$, which makes this result applicable also to light quarks.

Notice that the potentials $V_n(r)$ themselves do not change their
expressions as compared to Eq.~(\ref{Vseq}), only the ``mass"
factors are modified. In the limit of heavy quarks, the
Eichten--Feinberg expression (\ref{SO0}) is readily reproduced by
Eq.~(\ref{V02}) with the potentials given by Eq.~(\ref{Vseq}). In
the meantime, Eq.~(\ref{V02}) allows one to establish the
correction to the interaction (\ref{SO0}) which comes from the
string inertia. Indeed, in the limit $\bar m_i\gg\lan\nu\ran$, the
term $\tilde{\mu}$ in the denominators can be expanded so that
(for equal quark masses, $m_1=m_2=m$)
\be
V_{SD}(r)=V_{SD}^{(0)}(r)+\Delta V_{SD}(r),\quad \Delta V_{SD}(r)=\frac{\sigma^2}{6m^3}\veS\veL,
\ee where
$V_{SD}^{(0)}(r)$ is given by Eq.~(\ref{SO0}) and $\veS$ is the
total spin of the quark--antiquark pair. The term $\Delta
V_{SD}(r)$ has the meaning of the string correction for the
spin-dependent interquark potential, in analogy with the string
correction to the spin-independent interaction which comes from
the similar expansion of the last, angular-momentum-dependent term
in the Hamiltonian (\ref{Hm}) and which is discussed in detail in
the literature --- see, for example, Refs.~\cite{DKS,GD,westr}.
The correction $\Delta V_{SD}(r)$ constitutes a few MeV for the
lowest charmonium states.

The effect of the string inertia is more sizable for light quarks.
Indeed, in order to quantify the effect in this case let us
consider two limits: a heavy--light system, with $m_1=m$,
$m_2\to\infty$, and the limit of equal quark masses, $m_1=m_2=m$.
In the first limit $\mu_2\to\infty$, $\zeta_1\to 0$ and thus the
denominator reads (instead of simply $\mu_1^2$):
\be
\mu_1\left[\mu_1+\int^1_0\beta^2\nu(\beta)d\beta\right],
\label{den1}
\ee
whereas, for the light--light system, one has
$\mu_1=\mu_2=\mu_0$, $\zeta_i=1/2$, and the naive term $\mu_0^2$
in the denominator is substituted by the expression
\be
\mu_0\left[\mu_0+2\int^1_0\left(\beta-\frac12\right)^2\nu(\beta)d\beta\right].
\label{den2}
\ee
This is exactly the effect we are looking for: the string contributes the
total inertia of the system together with the quarks and this
string contribution is always present in the denominator, whenever
the angular momentum $\veL$ appears in the numerator. Due to this
modification the spin--orbital interaction is weaker than in the
``standard'' case when the moment inertia  of the rotating string
is neglected. For example, in heavy-light mesons one has
\be
\mu_1\cong\langle\sigma r\rangle,
\ee
so that, approximating $\nu(\beta)$ in Eq.~(\ref{den1}) by $\langle\sigma r\rangle$, one arrives at the effective
denominator change \cite{BNS2}
\be
\frac{1}{\mu_1^2}\to\frac34\frac{1}{\mu_1^2}.
\label{3400}
\ee

In the meantime, by an explicit calculation, one can find that, for the $P$-wave $B$-mesons,
the denominators change constitutes already about 50\%,
\be
\frac{1}{\mu_1^2}\to 0.52\frac{1}{\mu_1^2}.
\label{340}
\ee
Thus we conclude that, for light quarks and at low
angular momenta, the suppression of the spin--orbit interaction
due to the proper string inertia can reach 50\%.
Notice that this suppression has a purely dynamical origin.

\section{Discussion}\label{S4}

In this paper we studied in detail the spin-dependent interactions
in quarkonia. Recent lattice data for the spin-dependent
potentials in heavy quarkonia were analysed
using the theoretical formulae derived in the framework of
the FCM. The comparison with the lattice results appears to be in
good agreement for the potentials $V_0'(r)$, $V_2'(r)$, and
$V_3(r)$. In the meantime, for the potentials $V_1'(r)$ and
$V_4(r)$ (at short distances) a discrepancy between our
theoretical formulae and the lattice data is observed.

In case of the potential $V_1'(r)$ this discrepancy comes from the contribution of higher-order field correlators and
it may be related to the violation of the Gromes relation (\ref{Ge}) observed on the
lattice, since this relation is exact in the FCM for both
perturbative and nonperturbative parts of the potentials.

As far as the potential $V_4(r)$ is concerned, in the FCM the
potential $V_4(r)$ stems from the delta-function after smearing
with a finite vacuum correlation length, whereas on the lattice
this potential was found to have quite a different shape
--- it has a ``wrong" negative sign as compared to its ``standard"
OGE-inspired behaviour. We conclude therefore that this question
deserves a systematic study, primarily on the lattice. If the
behaviour of this potential reported in Ref.~\cite{Komas} is
confirmed, this will be a certain challenge for phenomenologists.

From the comparison of our analytic formulae with the lattice data
for the spin-dependent potentials in heavy quarkonia we extracted
the value of the gluonic correlation length. We showed that the
data were consistent with extremely small values of the vacuum
correlation length, less than 0.1 fm. This is an important result since,
on one hand, this result ensures a very small value of the parameter
which governs the cluster expansion in the stochastic QCD vacuum \cite{FCM}.
Indeed, this parameter can be calculated then as $\eta=\sigma T_g^2\approx 0.06\ll 1$ which ensures
a fast convergence of the series and justifies working in the Gaussian approximation for the field correlators, neglecting
higher correlators. On the other hand,
such a small vacuum correlation length justifies the use of the potential-type approaches to
quarkonia up to quite small interquark separations. In
particular, it validates the use of the QCD string approach. Thus
we stack to this potential-type approach in order to derive the
generalisation of the Eichten--Feinberg formula (\ref{SO0}) for
the case of light quarks. In particular, the pole quark masses in
the denominators are substituted by the averaged quark kinetic
energies and the string inertia contributes the total inertia of
the system and suppresses the angular-momentum-dependent
interactions. For example, for the spin--orbital interaction, the effect of the string inertia may lead to the
suppression by almost a factor of two, for low-lying states, while, for highly excited states, the spin--orbital interaction is suppressed
even more, by the factor $L^{-2/5}$ \cite{BNS2}.

\begin{figure}[t]
\begin{center}
\epsfig{file=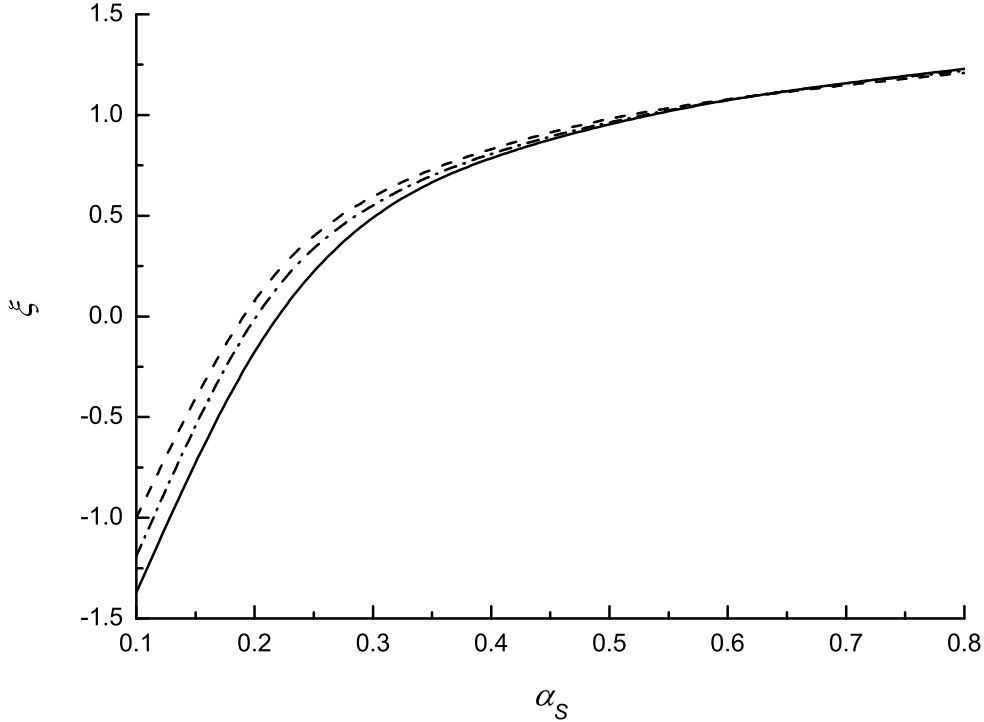,width=13cm}
\caption{The ratio $\xi=a_{SO}/t$ for the heavy--light quarkonium ($m_1=5$ GeV, $m_2=1$ MeV) as a function of the strong coupling constant
$\alpha_S$ for the string tension $\sigma=0.16$ GeV$^2$
(dashed line), $\sigma=0.19$ GeV$^2$ (dash-dotted line), and $\sigma=0.22$ GeV$^2$ (solid line).}\label{xifig}
\end{center}
\end{figure}

In conclusion we return to the quantity discussed in the very
beginning of the paper, namely to the ratio  of the spin-orbit and
tensor splittings $\xi=a_{SO}/t$. It was claimed in the Introduction that the experimental data on the
fine-structure splittings in heavy and heavy--light quarkonia appears to be
similar, in contrast to the predictions of the  heavy quark
models. The positive sign of $\xi$ and  its value $\xi\lesssim 1.0$
can be explained if sufficiently large (and confirmed by
phenomenology) values of the strong coupling constant are used and
also the correct definition of the dynamical quark masses are introduced.
Then, equipped with the generalised interquark spin-dependent
interactions in the form of Eq.~(\ref{V02}), as an example we
consider the behaviour of the ratio $\xi$  in the heavy--light
meson  with $m_1=m_Q=5$ GeV and $m_2=m_q=1$ MeV. The dependence
$\xi(\alpha_S)$ for three realistic values of the string tension
$\sigma$ is presented in Fig.~\ref{xifig}. From this figure one
can see that i) the dependence on the string tension is extremely
week; ii)  for the phenomenologically successful values of the
strong coupling constant $\alpha_S\simeq 0.5$ (see, for
example, Ref.~\cite{BB}), one has $\xi\simeq 1$, in agreement with
the experimental data, as given by Eq.~(\ref{4s}).

\begin{acknowledgments}
The authors would like to thank N. Brambilla and A. Vairo for useful discussions and Y. Koma for valuable comments and for providing the authors
with the raw lattice data.
This work was supported by the Federal Agen\-cy for Atomic Energy of Russian Fe\-de\-ration, by the Federal Programme of
the Russian Ministry of Industry, Science, and Technology No. 40.052.1.1.1112, and by the grant NSh-4961.2008.2 for the leading scientific
schools. A. M. B. and Yu. A. S. would like to acknowledge the financial support through the grants RFFI-06-02-17012 and RFFI-06-02-17120.
Work of A. N. was supported by RFFI-05-02-04012-NNIOa, DFG-436 RUS 113/820/0-1(R), PTDC/FIS/70843/2006-Fisica,
and by the non-profit ``Dynasty" foundation and ICFPM.
\end{acknowledgments}

\appendix

\section{Colour-electric contribution to the spin--orbit potential}\label{colel}

In order to simplify the notations we include the colour factors and the gauge coupling $g$ to the definition of the the gluonic field $A_\mu$.

We start from the single-quark Green's function and notice that
the spin-dependent contributions to the interquark interaction come from the Dirac projectors
$m-\hat{D}$ and from the combination $\sigma_{\mu\nu}F_{\mu\nu}$ in the exponent (see Eq.~(\ref{W1})),
\be
m-\hat{D}=\left(
\begin{array}{cc}
m-D_4 &i\vesig \veD\\
-i\vesig\veD & m+D_4
\end{array}
\right),
\quad
\sigma_{\mu\nu}F_{\mu\nu}=
\left(
\begin{array}{ll}
\vesig \veH& \vesig\veE\\
\vesig\veE&\vesig \veH
\end{array}
\right)\equiv h_H+\gamma_5h_E,
\label{mD}
\ee
where $h_E=\vesig\veE$, $h_H=\vesig\veH$, and for the future convenience we denote the spin-independent part of the interaction as $h_0$.
Now, in order to build the correction to the diagonal part of the Hamiltonian $h_0+h_H$ which comes from the off-diagonal part $h_E$, we are
to combine the latter with the off-diagonal part of the Dirac projector $m-\hat{D}$:
\be
{\rm Tr}(m_1-\hat{D}_1)\exp\left[-\int d\tau_1\gamma_4 (h_0+h_H+\gamma_5 h_E)\right].
\label{q1}
\ee
Expanding the exponent in Eq.~(\ref{q1}) and taking the Lorentz trace, one finds that the Hamiltonian has the form:
\be
H_1=h_0+h_H+\frac{\left(i\vesig_1\veD_1\right)}{\mu_1}h_E,
\ee
where the quark energy $\mu_1$ appeared from $D_4$. Notice that the Hamiltonian $H_1$ defines the propagation of the particle in the fifth, Schwinger
time $\tau_1$. Then, after the substitution $d\tau_1=dt/(2\mu_1)$, one finally arrives at the following colour-electric contribution to the
spin-dependent interaction:
\begin{eqnarray*}
\int\lan\left(i\vesig_1\veD_1\right)\left(\vesig_1\veE\right)W\ran\frac{dt}{2\mu_1^2}=
i\sigma_{1k}\sigma_{1i}\int\lan D_{1k}E_i W\ran\frac{dt}{2\mu_1^2}\Longrightarrow
\varepsilon_{ikl}\sigma_{1l}\int\frac{\partial}{\partial x_{1k}} \lan E_i W\ran\frac{dt}{2\mu_1^2}\\
\Longrightarrow\varepsilon_{ikl}\sigma_{1l}\int\dot{x}_{1k}\lan E_i(z)E_n (u)\ran ds_{n4}(u)\frac{dt}{\mu_1}
\Longrightarrow\frac{\vesig_1}{\mu_1}\int[\ver\times\dot{\vex}_1(t)]\lan E_xE_x\ran dt d\beta du_4\\
\Longrightarrow\vesig_1\frac{1-\zeta_1}{\mu_1}\int[\ver\times \dot{\ver}]_n\lan E_xE_x\ran dtdu_4 d\beta
\Longrightarrow\vesig_1\frac{1-\zeta_1}{\mu_1\tilde{\mu}}\int[\ver\times \vep]\lan E_xE_x\ran dt d\beta du_4 \\
=(\vesig_1\veL)\frac{1-\zeta_1}{\mu_1\tilde{\mu}r}\int_0^\infty d\nu\int^r_0 d\lambda \left[D^E + D^E_1+
(\lambda^2+\nu^2)\frac{\partial D_1^E}{\partial \nu^2}\right]dt=\int V_{1SO}^E(r) dt.
\end{eqnarray*}
Supplying the last equation with a similar expression for the second quark, one arrives at
\be
V_{SO}^{E}(r)=\left[(1-\zeta_1)\frac{\vesig_1\veL}{4\mu_1\tilde{\mu}}+(1-\zeta_2)\frac{\vesig_2\veL}{4\mu_2\tilde{\mu}}\right]\frac1r\frac{dV_0'}{dr},
\ee
with
\be
\frac1r\frac{dV_0'}{dr}=\frac2r\int_0^\infty d\nu\int^r_0 d\lambda\left[D^E+D^E_1+(\lambda^2+\nu^2)\frac{\partial D_1^E}{\partial\nu^2}\right],
\ee
in agreement with Eq.~(\ref{V02}).

For heavy quarks,
\be
\mu_i\approx m_i,\; (i=1,2),\quad \tilde{\mu}\approx\mu\approx \frac{m_1m_2}{m_1+m_2},\quad \zeta_1\approx\frac{m_1}{m_1+m_2},
\ee
and thus one arrives at the corresponding term in Eq.~(\ref{SO0}).

Note that $V_0'(r)$ appears only in the first order of this expansion and cannot be
exponentiated --- in contrast to $V'_1(r)$ and $V'_2(r)$ --- see Appendix~\ref{colmagn}.
This implies that the Gromes relations can only be applied to the first order
expansion.

\section{Color-magnetic contribution to the spin-orbit potential}\label{colmagn}

We now turn to the colour-magnetic contributions to the spin--orbital interaction. As before, we include the colour factors as well as the gauge
coupling to the definition of the gluonic field $A_\mu$.

Obviously, we consider now the diagonal terms in the matrix (\ref{3}). Then one finds from Eq.~(\ref{W1}):
\beq
L_{SO}^{(H)}&=&-\int V_{SO}^H(r)dt=\int ds_{ik}(w)d\tau_1\sigma^{(1)}_{\alpha\beta}\lan F_{ik}(w)F_{\alpha\beta}(x_1(\tau_1))\ran+(1\to 2)\nonumber\\[-2mm]
\label{LSOH}\\[-2mm]
&=&\int \frac{dt}{2\mu_1}\int^1_0 d\beta\int d|\veu_1|\frac{\beta-\zeta_1}{\tilde{\mu}}\;\sigma_{1n}L_m\lan H_m(w)H_n(x_1)\ran+(1\to 2),\nonumber
\eeq
where Eqs.~(\ref{19a}) and (\ref{a250}) were used and $d\tau_i=dt/(2\mu_i)$, $u_{i\mu}=w_\mu(\beta,t)-x_{i\mu}(t)$ ($i=1,2$).
Then, using Eq.~(\ref{Hs0}) for the
correlator of the magnetic fields $\lan H_mH_n\ran$, one can find the magnetic contribution to the spin--orbital potential in the form:
\beq
V_{SO}^{H}(r)=-\int_0^\infty d\nu\int^1_0 d\beta\left[(\beta-\zeta_1)\left(D^H+D_1^H+\lambda^2\frac{\partial D_1^H}{\partial\lambda^2}\right)\right.
\nonumber\\[-2mm]
\label{VSOH1}\\[-2mm]
\left.+(1-\zeta_1)\nu^2\frac{\partial D_1^H}{\partial\nu^2}\right]\frac{\vesig_1\veL}{2\mu_1\tilde{\mu}}+(1\to 2)\nonumber.
\eeq
The two terms in Eq.~(\ref{VSOH1}) can be combined together with the help of the variable change $\beta\to 1-\beta$ and,
introducing the variable $\lambda=\beta r$, one can arrives at the final formula:
\beq
V_{SO}^{(H)}(r)=-\frac{1}{2r}\int_0^\infty d\nu\int_0^r d\lambda\left[\left(1-\frac{\lambda}{r}-\zeta_1\right)\left(D^H
+D_1^H+ \lambda^2 \frac{\partial D^H_1}{\partial\lambda^2}\right)\right.\nonumber\\[-2mm]
\label{VSOH2}\\[-2mm]
\left.+(1-\zeta_1)\nu^2\frac{\partial D_1^H}{\partial\nu^2}\right]
\left(\frac{\vesig_1\veL}{\mu_1\tilde{\mu}}+\frac{\vesig_2\veL}{\mu_2\tilde{\mu}}\right),\nonumber
\eeq
which is equivalent to the sum of the potentials $V_1'(r)$ and $V_2'(r)$ --- see Eq.~(\ref{Vseq}).

\end{document}